\documentclass{aa}  

\usepackage{graphicx}
\usepackage{verbatim}
\usepackage{txfonts}
\usepackage{multirow}
\usepackage{natbib}
\begin{document}

   \title{Non-Local Thermodynamic Equilibrium Transmission Spectrum Modelling of HD209458b}

   \author{M. E. Young
          \inst{1}
          \and
          L. Fossati
          \inst{1}
          \and
          T. T. Koskinen
          \inst{2}
          \and
          M. Salz
          \inst{3}
          \and
          P. E. Cubillos
          \inst{1}
          \and
          K. France
          \inst{4}
          }

   \institute{Space Research Institute (IWF), Austrian Academy of Science,
              Schmiedlstra{\ss}e 6, 8042 Graz\\
              \email{mitchelleric.young@oeaw.ac.at}
         \and
             Lunar and Planetary Laboratory, University of Arizona, 1629 E. University Blvd., 85721 Tuscon
         \and
             Hamburg Observatory, University of Hamburg, Gojenbergsweg 112, 21029 Hamburg
         \and
             Laboratory for Atmospheric and Space Physics, University of Colorado Boulder, 80309 Boulder
             }

   \date{}

  \abstract
   {Exoplanetary upper atmospheres are low density environments where radiative processes can compete with collisional ones and introduce non-local thermodynamic equilibrium (NLTE) effects into transmission spectra.}
   {We develop a NLTE radiative transfer framework capable of modelling exoplanetary transmission spectra over a wide range of planetary properties.}
   {We adapt the NLTE spectral synthesis code Cloudy to produce an atmospheric structure and atomic transmission spectrum in both NLTE and local thermodynamic equilibrium (LTE) for the hot Jupiter HD\,209458b, given a published $T$--$P$ profile and assuming solar metallicity. Selected spectral features, including H$\alpha$, Na{\sc i} D, He{\sc i} $\lambda$10830, Fe{\sc i} \& {\sc ii} ultra-violet (UV) bands, and C, O and Si UV lines, are compared with literature observations and models where available. The strength of NLTE effects are measured for individual spectral lines to identify which features are most strongly affected.}
   {The developed modelling framework computing NLTE synthetic spectra reproduces literature results for the He{\sc i} $\lambda$10830 triplet, the Na{\sc i} D lines, and the forest of Fe{\sc i} lines in the optical. Individual spectral lines in the NLTE spectrum exhibit up to 40\% stronger absorption relative to the LTE spectrum.}
   {}

   \keywords{planets and satellites: general --
                planets and satellites: individual: HD\,209458b --
                planets and satellites: atmospheres --
                radiative transfer -- 
                techniques: spectroscopic
               }

   \maketitle
%

\section{Introduction}

    Over the course of the last two decades, astronomers have observed measurable reductions in the observed light of stars as exoplanets pass in front of them \citep{charbonneau00, henry00, mazeh00}. Planets that transit their host stars represent an extremely important phenomenon because they uniquely provide critical characteristics of planetary systems, such as the ratio of the planetary and stellar radii and the ratio of the planetary orbit and the stellar radius, as well as the opportunity to characterize the planetary atmosphere. 
    
    The first exoplanetary atmosphere observed in transmission was detected by \citet{charbonneau02}, who found sodium absorption in the atmosphere of HD\,209458b. In the following years there have been additional detections in exoplanetary atmospheres of chemical elements such as sodium \citep[e.g.,][]{redfield08, snellen08, wood11}, potassium \citep[e.g.,][]{colon12, sing11}, magnesium \citep[e.g.,][]{fossati10,sing19}, hydrogen \citep[e.g.,][]{vidal-madjar03, lecavelier10, yan18}, carbon and oxygen \citep[e.g.,][]{vidal-madjar04}, silicon \citep[e.g.,][]{linsky10, schlawin10}, helium \citep[e.g.,][]{nortmann18,salz18,alonso-floriano19}, iron \citep[e.g.,][]{haswell12,sing19,cubillos2020}, calcium \citep[e.g.,][]{yan19,turner2020}, and molecules including H$_2$O \citep[e.g.,][]{barman07, deming13, fraine14, mccullough14, kreidberg15, sanchez-lopez19}, CO \citep[e.g.,][]{snellen10, rodler13, brogi16}, and possibly TiO \citep[e.g.,][]{nugroho17, espinoza19}.
    
    
    All atmospheric transit detections are identified through wavelength dependent variations in apparent planetary radius, observed using either narrow-band photometry or high-/low-resolution spectroscopy, as follows. With information about the absorbing species, such as the column density at which the material becomes optically thick to given wavelengths, this can be translated into the altitude of the absorbing material. In some cases, the inferred altitude of the absorbing material implies the presence of planetary gas beyond the Roche lobe, demonstrating that the planet undergoes atmospheric escape. Escaping hydrogen has been detected in this fashion for HD\,209458b \citep{vidal-madjar03}, HD\,189733b \citep{lecavelier10}, GJ\,436b \citep{ehrenreich15}, GJ\,3470b \citep{bourrier18}, K2-18b \citep{dossantos2020}, while escaping heavy atoms have been detected for HD209458b \citep{vidal-madjar04,linsky10,cubillos2020}, WASP-12b \citep{fossati10,haswell12} and WASP-121b \citep{salz19,sing19}. Of these planets, HD\,209458b is by far the most well studied.
    
    The difficulty in modelling transmission spectra for planets with escaping atmospheres is that while the assumption of local thermodynamic equilibrium (LTE) holds in regions of high density gas, namely the lower atmosphere, the assumption breaks down in the low density upper atmosphere, particularly in the exosphere where excitation and ionization processes are not collisionally dominated. In the upper atmosphere, where the radiation field (ie. the incident stellar radiation) is not representative of the local gas temperature, ionization stages and level populations are best described by non-local thermodynamic equilibrium (NLTE) distributions. For example, the effects of NLTE and photo-ionization have already been explored for sodium in the atmosphere of HD209458b \citep{barman02,fortney03}. These effects are not detectable at low spectral resolution probing the lower and middle atmosphere but they have significant consequences for the interpretation of high resolution absorption spectra probing the upper atmosphere \citep{lavvas14}.  More recent investigation of HAT-P-1b, HAT-P-12b, HD189733b, WASP-69b, WASP-17b, and WASP-39b were unable to distinguish between LTE and NLTE interpretations of the sodium doublet \citep{fisher19}, also implying the need for high resolution observations to identify the effects of NLTE in escaping atmospheres.   
    
    
    In this work, we present a framework for simulating NLTE atomic transmission spectra of exoplanets for any set of given planetary and system parameters. Our framework is generalized to produce spectra at any desired spectral resolution in any wavelength band, as long as the appropriate line lists and rate coefficients are available. Here, we apply this framework to model the ultraviolet to near-infrared transmission spectrum of HD\,209458b in both LTE and NLTE, and compare features of the resultant spectra to observations available in the literature.

\section{Transmission Spectrum Modelling}
\label{sec:trans_spec_mod}

    To simulate exoplanetary transmission spectra, we adapt the NLTE spectral synthesis and plasma simulation code Cloudy v17.01 \citep{ferland17}, which is designed to simulate conditions in interstellar matter under a broad range of conditions and predict their spectra. Cloudy simultaneously and self-consistently calculates ionization, chemistry and radiative transfer, with a chemical network that covers atoms, ions and molecules for elements up to and including Zn. 
    
    In this work, we use a temperature--pressure ($T$--$P$) profile calculated from the model of \citet{koskinen13}, as in \citet{lavvas14}. While Cloudy has the capability to calculate the temperature profile, the thermal structure calculations are complicated and Cloudy has not been properly tested over such a wide range of pressures in close-in exoplanetary atmospheres. In addition, Cloudy does not include hydrodynamic escape that shapes the temperature profile and the chemical composition in the upper atmosphere. Our current aim, instead of trying to calculate the temperature profile, is to develop a framework that allows us to test the influence of different temperature profiles on the observations.
    
    
    At minimum, then, our setup with Cloudy requires an atmospheric temperature profile, the total atmospheric density profile of hydrogen (including atoms, ions and molecules), orbital separation, and both the bolometric luminosity and SED of the host star as input. Further information, such as additional elemental density profiles, atmospheric microturbulence, transit impact parameter, orbital period, system age, and stellar effective temperature can be given to fit observed transmission spectra, but are not required as model inputs. By default, Cloudy does not include stratification of the atmosphere by gravity and we developed a simple scheme to incorporate this in our simulations (Section~\ref{subsc:atm_mod}).
    
    This work represents the continuation of adapting Cloudy to exoplanetary atmospheres (see \citet{salz15} for the initial work). The benefit of this development is that the model can calculate a wide range of ionization states and account for NLTE level populations in transit depth and radiative transfer calculations. At this point, however, our modeling scheme makes several simplifying assumptions:
    \begin{enumerate}
        \item The atmospheric model is 1-D. In reality, planetary atmospheres display differences in temperature and densities between the equator and the poles, as well as between the day and night sides. On some ultra-hot Jupiters (UHJs) these differences can reach values large enough to promote formation and dissociation of molecules \citep{parmentier18}, leading to differences in chemistry as well. We note that our modeling framework with Cloudy is flexible in the sense that we could calculate transit depths based on output from 3-D models, although we have not followed that approach here.
        \item Both the stellar and planetary atmospheres are assumed to be static and perfectly spherical in shape. In terms of the planetary atmosphere, no additional broadening mechanisms for spectral features (i.e. planetary rotation, winds, turbulence, pressure, etc.) other than thermal and natural broadening are included. 
        \item In terms of the stellar atmosphere, we do not consider the effects of stellar activity or convection in our transmission spectra. We ignore any effect stellar activity may have on our transmission spectra and light curves, but note that it is possible to add the inclusion of stellar spots by replacing part of the stellar disc in the calculations described in Section \ref{sec:trans_spec} with one or more additional stellar SEDs of different temperatures (i.e., those of the spots). The effects of individual convection cells are considered to be inconsequential on the spatial scale of our models, and when using an observed stellar spectrum for the input SED, the average effect of granulation is included.
        \item While HD\,209458 is not rotating rapidly enough to deform its spherical shape, and therefore is unaffected by this limitation, our current methodology does not make considerations for other stars that may be rapid rotators. Rapid stellar rotation can lead to both deformation of the spherical shape of a star and gravity darkening that may affect a transit light curve if there is a spin-orbit misalignment in the system \citep{masuda15, barnes09, ahlers15}.
        \item We do not include atmospheric molecules (other than H$_2$) or clouds and aerosols in our simulations of the transmission spectrum. Cloudy calculates ionization up to very high ionization states and has the capability to simulate some chemistry, including the neutral and ionized molecular species listed in the Appendix of \citet{ferland17}. The code is well suited and adaptable to simulating interstellar clouds and exoplanetary upper atmospheres, but its capability for simulating the complex chemistry of exoplanetary middle and lower atmospheres is more limited and has not been adapted to include, for example, aerosol condensation.
        
        \item Finally, while we simulate the effects of stellar limb-darkening in our transmission spectra as described in Section~\ref{sec:trans_spec}, we ignore the effect of chromospheric/coronal limb-brightening in individual spectral lines. The synthetic spectral model we fit to generate our limb darkening curves does not account for chromospheric/coronal contributions, but our setup is capable of calculating limb-brightening effects if a suitable spectral model is used.
    \end{enumerate}
    
    Currently, the inclusion of any of the following would be an improvement on our modelling scheme: hydrodynamics, molecular chemistry, pressure broadening, 3D atmospheric structure (day/night or equator/pole variations), stellar activity, and Doppler effects (planetary and stellar rotation, orbital motion, etc). Cloudy in its current form does not include hydrodynamics or pressure broadening and cannot compute 3D models. While all three may be approximated in some form, it is beyond the capabilities of our current modelling scheme to do so.
    
    It would be possible to include both Doppler effects and stellar activity in our scheme by making additional assumptions, and with additional resources. For the full inclusion of Doppler effects, knowledge of the planetary and stellar rotation rates, inclinations of the rotational axes with respect to the orbital plane, and atmospheric circulation on the planet would need to be assumed. Currently, transmission spectra are modelled at mid transit, where only the planetary and stellar rotational rates affect the spectrum by broadening the lines. Spectral line broadening by stellar rotation is already included in the observed solar spectrum used as the illumination source.
    
    Stellar activity could be approximated by making a patchwork stellar disc of multiple observed spectra, instead of a smooth single spectrum disc. This would require making assumptions concerning the number and size of stellar spots, differences in the spectra of the radiation emitted from the spots, and the time dependence of the formation and dissolution of spots. It would also require multiple limb darkening models to be able to limb darken the individual spots. While the framework can be modified to include all of this, any effects of stellar activity apparent in the transmission spectrum would likely be rivalled by the uncertainties arising from the large number of assumptions.

    \subsection{HD\,209458b Atmospheric Model}
    \label{subsc:atm_mod}
    
        In this work, we use the same $T$--$P$ profile at the substellar point as \citet{lavvas14} in the lower atmosphere, originally taken from \citet{showman09}. In the upper atmosphere, we use a $T$--$P$ profile calculated with the model of \citet{koskinen13}, which includes effects of atmospheric escape that are not present in the \citeauthor{lavvas14} profile. These two $T$--$P$ profiles are joined continuously at 1 $\mu$bar. 
        
        We choose the substellar temperature profile rather than the terminator profile for several reasons. In our setup, Cloudy’s operation is limited to calculating radiative transfer along a path that is parallel to the incident stellar radiation. With a terminator profile necessarily being orthogonal to the radiation, there is no component of the radiation that travels parallel to this profile. The substellar temperature profile maintains self-consistency with the resultant substellar elemental abundance profiles. Additionally, other assumptions made in the modelling scheme or the uncertainty in the modelled T-P profile will likely have a larger impact on the resultant transmission spectrum than the sampled location of the $T$--$P$ profile.
        
        The literature $T$--$P$ profile we employed is sampled at a total of 747 points throughout the atmosphere, 549 in the upper atmosphere and 198 in the lower atmosphere. In the lower atmosphere, these sampling points are evenly spaced in log $P$ space, but in the upper atmosphere they are distributed such that there is higher log $P$ resolution directly above the 1 $\mu$bar level, and less resolution higher in the atmosphere, causing an abrupt change in the sampling rate at 1 $\mu$bar. We re-sample the $T$--$P$ profile at 200 points evenly distributed in log $P$ space, maintaining a constant sampling rate throughout the atmosphere.
        
\begin{figure*}
        \begin{center}
        	\includegraphics[scale=0.9]{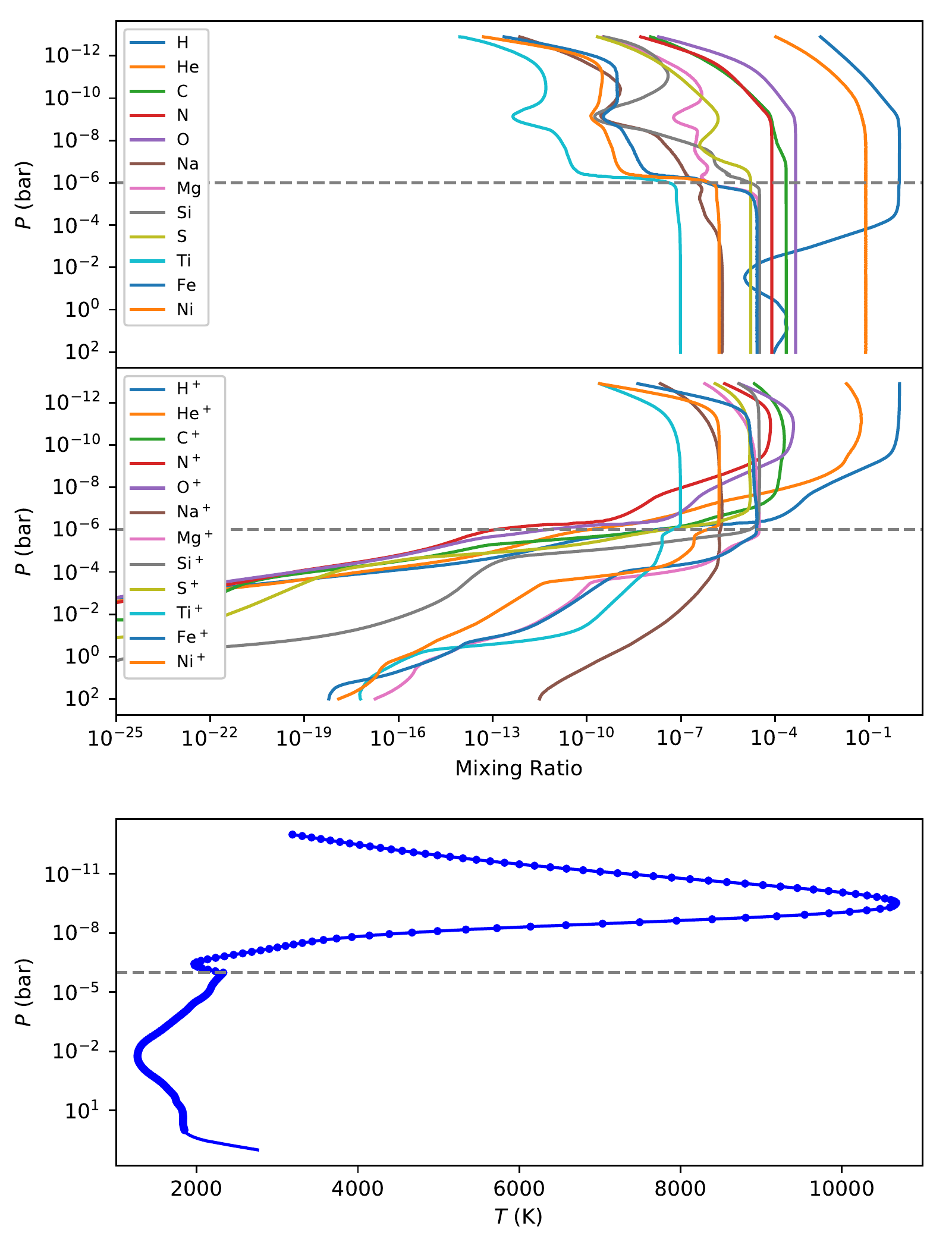}
            \caption{HD\,209458b 1-D atmospheric model mixing ratios (top) and temperature profile (bottom). Mixing ratios are day-side averages calculated by Cloudy. The temperature profile is taken from \citet{lavvas14} (lower) and \citet{koskinen13} (upper), joined at 1 $\mu$bar (dashed gray line). The dots indicate the 200 points to which the $T$--$P$ profile has been re-sampled.}
            \label{fig:profiles}
        \end{center}
        \end{figure*}

\begin{figure*}
        \begin{center}
        	\includegraphics[scale=0.9]{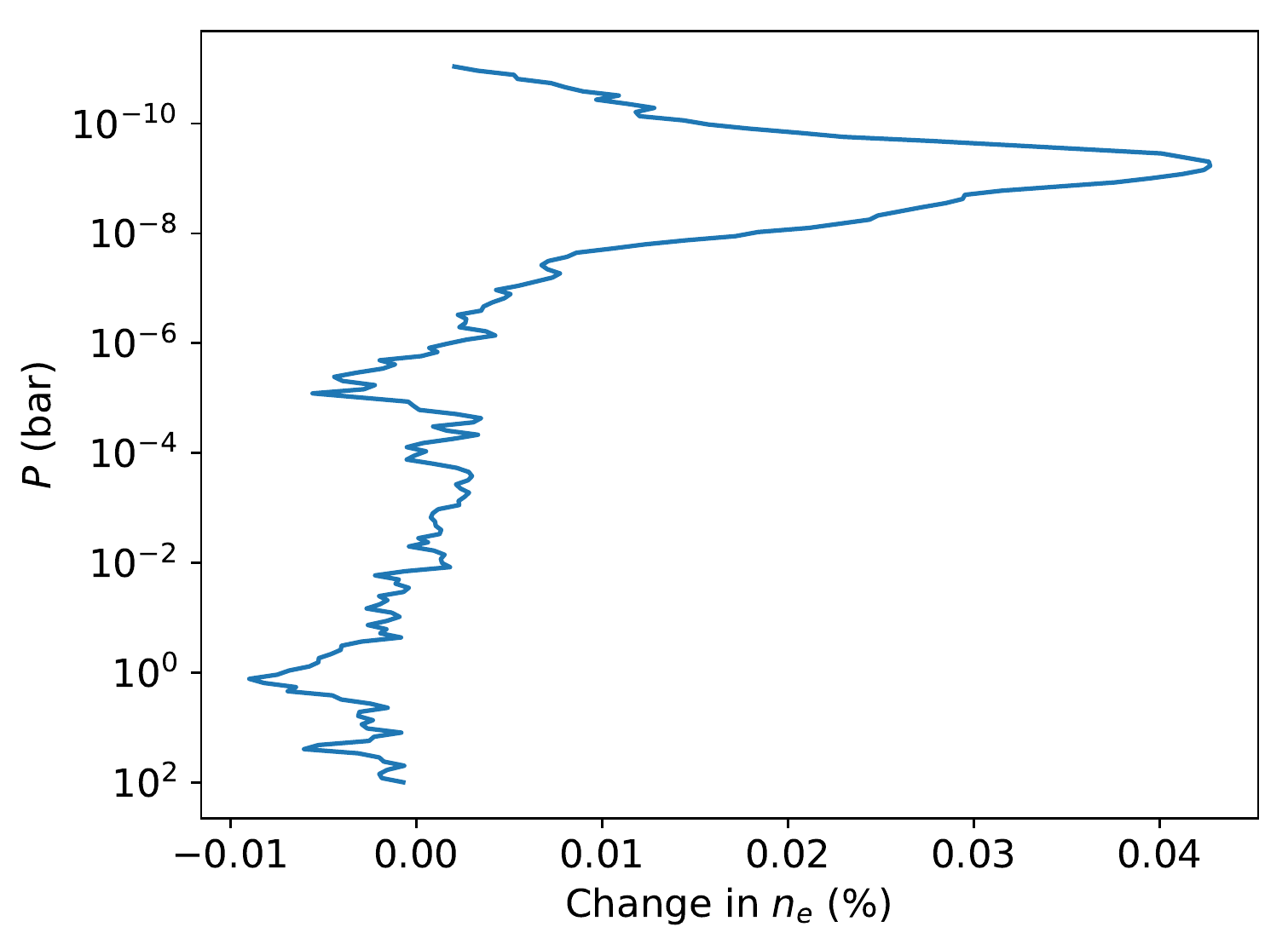}
            \caption{Difference in modeled e$^-$ density, presented as the running mean over 10 sampling points, for models that do and do not include N, S, Ti, and Ni. Models are otherwise equivalent.}
            \label{fig:e_density}
        \end{center}
        \end{figure*}

        We include H, He, C, N, O, Na, Mg, Si, S, Ti, Fe and Ni in the Cloudy simulation with solar abundance ratios, and compute the composition and ionization structure of the atmosphere at the sub-stellar point. The quiet Sun irradiance reference spectrum by \citet{woods09} is used to illuminate the planet in place of an HD\,209458 SED, scaled the to half the stellar luminosity to approximate integration over the planetary day side. The altitude scale of the atmosphere is determined iteratively as the composition evolves, according to 
        \begin{equation}
            r_i = \left( \frac{1}{r_{i-1}}+\frac{kT_i}{GMm_i}\ln{\frac{P_i}{P_{i-1}}} \right)^{-1}
        \end{equation}
        where $r_i$ is the radius of the $i$th layer (measured from the center of the planet), $k$ is the Boltzmann constant, $T_i$ is the temperature of the $i$th layer, $G$ is the gravitational constant, $M$ is the planetary mass, $m_i$ is the mean molecular weight of particles at the $i$th layer, and $P_i$ is the pressure of the $i$th layer. This process is iterated upon until the ionization and altitude structures of the atmosphere change by less than $1\%$ between iterations at all sampling points. Because Cloudy lacks the facility to include both hydrodynamic effects and our framework doesn't account for the Roche potential, we limit the atmosphere to a maximum radius of $1\,R_\star$ (occurring at a pressure of approximately $2\times10^{-12}\,$bar under the initial conditions). Outside of this radius, the atmospheric scale height diverges. Planetary and system parameters for HD\,209458b are listed in Table~\ref{tab:209_param}, and the final density profiles for individual species, as well as the input temperature profile, are presented in Fig.~\ref{fig:profiles}. As a demonstration of the impact that adding additional elemental species into the model has on the atmospheric structure, Fig.~\ref{fig:e_density} presents the difference in the e$^-$ density for models with and without N, S, Ti, and Ni.
        
        Note again that our model assumes hydrostatic equilibrium because Cloudy does not account for atmospheric escape, and we do not include the Roche potential due to stellar gravity. These assumptions are roughly valid below the sonic point and/or the Roche lobe boundary but invalid at higher altitudes. Our results therefore represent upper limits for atomic line core absorption that probe regions outside of the Roche lobe. The coupling of Cloudy to a model of atmospheric dynamics is subject to future work that will quantify the impact of escape and NLTE effects on detectable signatures.
        
        \begin{table}
        	\centering
        	\caption{HD\,209458b Planetary and System Parameters.}
        	\label{tab:209_param}
        	\begin{tabular}{lcc}
        		\hline
        		\noalign{\smallskip}
        		Parameter & Value & Reference\\
        		\noalign{\smallskip}
        		\hline
        		\noalign{\smallskip}
        		$T$ (days) & $3.52474859\pm0.00000038$ & 1\\
        		$b$ & $0.499\pm0.008$ & 2\\
        		Orbital & \multirow{2}{*}{$0.0488\pm0.0093$} & \multirow{2}{*}{1}\\
        		Separation (AU) &  & \\
        		Transit Depth (\%) & $1.5000\pm0.0024$ & 1\\
        		$P_0$ (bar) & 0.01 & 1\\
        		$R_{\rm P}$ ($R_\oplus$) & $15.6\pm0.2$\phantom{0} & 1\\
                $R_*$ ($R_\odot$) & $1.19\pm0.02$ & 1\\
                $T_\mathrm{eff}$ (K) & $6091\pm10$\phantom{00} & 1\\
                $L_*$ ($L_\odot$) & $1.77\phantom{00}^{+0.13}_{-0.14}$ & 3\\
        		\noalign{\smallskip}
        		\hline
        	\end{tabular}
        	\tablebib{(1)~\citet{stassun17}; (2) \citet{evans15}; (3) \citet{delburgo16}.}
        \end{table}

    \subsection{Transmission Spectrum}
    \label{sec:trans_spec}
        
        To obtain the transmission spectrum, we use Cloudy to compute the line-of-sight absorption and scattering of the stellar spectral energy distribution (SED) through the limb of the atmosphere at different altitudes. Our assumption of a spherically symmetric, static planetary atmosphere, the properties of which vary only with altitude, requires only a single Cloudy simulation to be computed at each impact parameter from the center of the planet. By integrating these simulations around the line-of-sight axis in the azimuthal direction, this produces a series of concentric rings with constant transmission over any given ring, reducing the total number of simulations necessary to model the full transmission spectrum.
        
\begin{figure}
        	\includegraphics[width=\columnwidth]{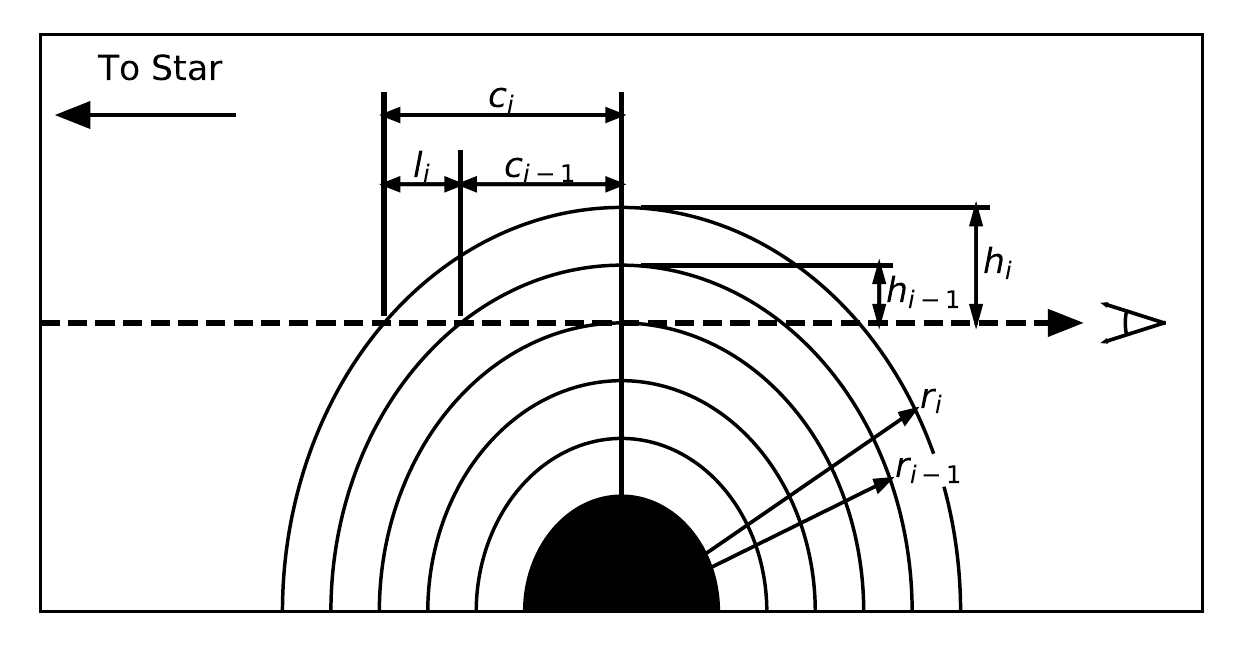}
            \caption{Schematic diagram of transmission chords along the line-of-sight for transmission spectrum calculation. $c$ is half the chord length along line of sight, $l$ is the path length through a given layer of atmosphere, $h$ is the height down from the layer to the altitude of the transmission chord at the terminator, $r$ is the radius of the layer, and the subscripts denote which layer.} 
            \label{fig:schematic}
        \end{figure}

        The 1-D model properties are mapped onto concentric spherical shells, and path lengths through successive layers of atmosphere along line-of-sight chords are calculated as
        \begin{equation}
        l_i = c_i - c_{i-1} = \sqrt{2r_ih_i-h_i^2} - \sqrt{2r_{i-1}h_{i-1}-h_{i-1}^2}
        \end{equation}
        where $l_i$ is the path length through the $i$th layer, $c_i$ is half the chord length along line-of-sight through the $i$th layer, $r_i$ is the radius of the $i$th layer, and $h_i$ is the height down from the $i$th layer of the atmosphere to the chord altitude at the terminator. A schematic diagram of this geometry is displayed in Fig.~\ref{fig:schematic}. These lengths, along with the atmospheric properties of their respective layers, are stacked and entered into Cloudy, using the Cloudy $table$ commands, as the line-of-sight transmission medium. 
        
        Cloudy computes output spectra spanning wavelength from 29.98~m (10~MHz) to 12.40~fm (100~MeV), at a default spectral resolution of $R=300$ for $\lambda>$1.01~\AA\ ($<$12.24~keV), and $R=33.333333$ otherwise. We increase the output spectral resolution to $R=100\,000$ over a wavelength range of $920\,$\AA$\,<\lambda<11000\,$\AA, while maintaining the default resolutions outside of this range. We select these high-resolution boundaries to include a large number of Lyman series lines at the short wavelength end, and the He{\sc i} $\lambda10830$ triplet at the long end. While the commonly accepted convention in spectroscopy is to use vacuum wavelengths for wavelengths shorter than $2000\,$\AA, and air wavelengths for wavelengths longer than $2000\,$\AA, Cloudy outputs the full spectrum in vacuum wavelengths to avoid a discontinuity in the spectrum at $2000\,$\AA. The output from the individual Cloudy transmission simulations is converted into the transmission spectrum by summing the contributions of the individual spectra, weighted by the relative area of the stellar disk they cover and the respective limb darkening. 
        
        We use a non-linear relation for stellar limb-darkening \citep{claret00}, of the form 
        \begin{equation}\label{eqn:limbdark}
            f_\mu = \frac{I(\mu)}{I(\mu=1)} = 1 - c_0(1-\mu^{0.5}) - c_1(1-\mu) - c_2(1-\mu^{1.5}) - c_3(1-\mu^2)
        \end{equation}
        where $I(\mu)$ is the line-of-sight intensity at position $\mu=\cos\theta$ ($\theta$ is the angle between the line-of-sight and the emergent intensity), and $c_0$, $c_1$, $c_2$, and $c_3$ are the limb-darkening coefficients \citep{salz19}. The coefficients are derived by fitting stellar limb-darkening curves from a limb-darkened PHOENIX model with an effective temperature of $T_\mathrm{eff}=6100\,K$, surface gravity of $log\,g=4.5$, and solar metallicity \citep{husser13}. We note that this PHOENIX model does not include computation of the stellar chromosphere or transition region, and is therefore incorrect in the estimation of centre-to-limb variations for emission lines forming in the stellar chromosphere and/or corona. The lack of chromospheric and coronal stellar emission also makes the PHOENIX model unsuitable for use as the central illumination source for the transmission spectrum because of the missing input high-energy stellar photons affecting the planetary atmospheric composition and of the missing UV stellar emission lines. Conversely, observed solar limb spectra do not exist at high resolution for the full wavelength range considered in this work, making them unsuitable for limb-darkening calculations.
        
\begin{figure}
        	\includegraphics[width=\columnwidth]{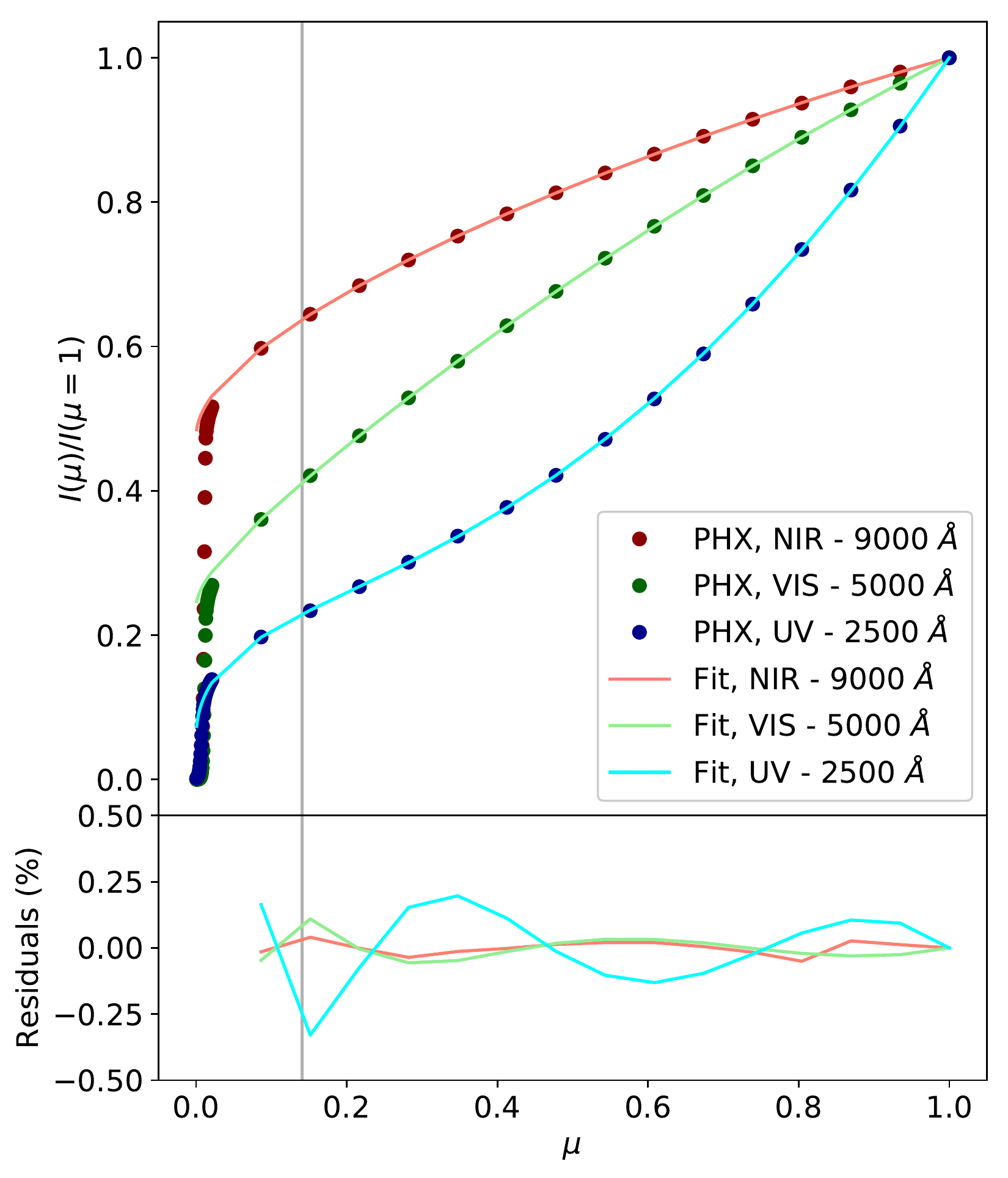}
            \caption{PHOENIX model stellar limb darkening curves (circles) and non-linear fits (lines), evaluated for three sample wavelengths, evaluated in 1 \AA~windows. Red:\,9000\,\AA, Green:\,5000\,\AA, and Blue:\,2500\,\AA. The vertical grey line represents the largest radius we evaluate limb darkening at, 0.99~$R_\star$., equivalent to $\mu\approx0.14$.} 
            \label{fig:LD_fits}
        \end{figure}

        The wavelength dependence of the limb darkening relation is accounted for by evaluating Eqn. \ref{eqn:limbdark} every $1\,$\AA\  (i.e., the sampling of the PHOENIX spectra) between $920<\lambda<11000\,$\AA. Example fits and residuals are presented in Fig.~\ref{fig:LD_fits} for three different limb darkening curves, one in the near-infrared (NIR), one in the visible (VIS), and one in the mid-ultraviolet (MUV).  
        
\begin{figure}
        	\includegraphics[width=\columnwidth]{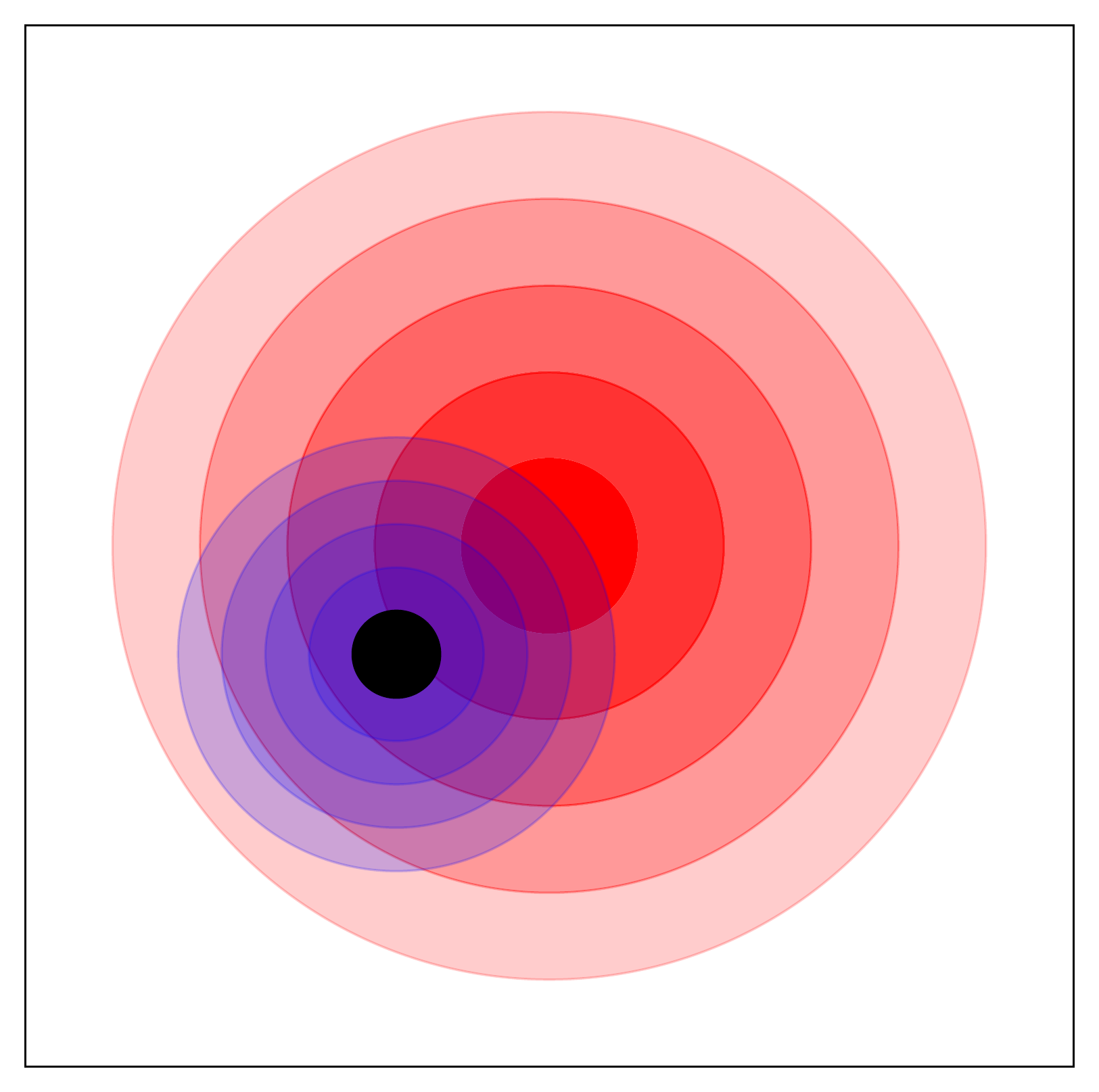}
            \caption{Schematic diagram of the line-of-sight view of a transiting planet in our framework. The shaded red circles are the stellar rings of discrete limb darkening, the shaded blue circles are different layers of the planetary atmosphere, and the black circle is the bulk disc of the planet.}
            \label{fig:discs}
        \end{figure}

        The stellar disc is divided into rings, similar to the planetary atmosphere, every 0.01~$R_\star$ out to a maximum of 0.99~$R_\star$, and the limb darkening is evaluated independently for each ring. To generate the summation weights for building the transmission spectrum, we calculate the overlapping areas of the planetary and stellar rings using the formula for intersecting circles, 
        \begin{multline}
            A(r,R) = r^2\cos^{-1}\left[\frac{D^2+r^2-R^2}{2Dr}\right]+R^2\cos^{-1}\left[\frac{D^2+R^2-r^2}{2DR}\right] \\
            -\frac{1}{2}\sqrt{(-D+r+R)(D+r-R)(D-r+R)(D+r+R)}\,,
        \end{multline}
        where $A(r,R)$ is the area of the intersection, $r$ and $R$ are the radii of the two circles, respectively, and $D$ is the distance between the centers of the circles. To convert these circular intersections to ring intersections, we subtract off the circular intersections of the next smaller ring radii of each the planetary and stellar rings, 
        \begin{equation}
            A_{ij} = A(r_i,R_j) - A(r_{i-1},R_j) - A(r_i,R_{j-1}) + A(r_{i-1},R_{j-1})\,,
        \end{equation}
        where $A_{ij}$ is the overlapping area of planetary ring $i$ and stellar ring $j$, $r_i$ is the radius of the $i$th planetary ring, and $R_j$ is the radius of the $j$th stellar ring. An illustrative schematic of a planet in-transit in this framework is presented in Fig.~\ref{fig:discs}. 
        
        We then take the summation weights to be 
        \begin{equation}
            w_{ij} = \frac{A_{ij}}{A_\star}\cdot\frac{I(\mu_j,\lambda)}{I(\mu=1,\lambda)}\,,
        \end{equation}
        where $w_{ij}$ is the weight of the overlapping $i$th and $j$th rings, $A_{ij}/A_\star$ is the overlap area of the $i$th and $j$th rings relative to the area of the stellar disc, and $I(\mu_j,\lambda)/I(\mu=1,\lambda)$ is the wavelength dependent limb darkening of the $j$th ring. For building the relative flux transit spectrum out of the individual ring spectra, we assume that the distance between the centers of the stellar and planetary discs is the impact parameter, $b$ (i.e., the planet is at mid transit), and the summation is performed according to
        \begin{equation}
            \frac{F_\lambda}{F_{\lambda,\star}} = \frac{\sum_{i,j}F_{\lambda,i}\cdot w_{ij}}{F_{\lambda,\star}}\,,
        \end{equation}
        where $F_\lambda$ is the in transit flux spectrum, $F_{\lambda,\star}$ is the out of transit stellar flux spectrum at disc centre, and $F_{\lambda,i}$ is the flux spectrum of the $i$th planetary ring. The quiet Sun irradiance reference spectrum is a disc integrated stellar spectrum, and therefore already includes the average effect of stellar limb darkening. We extract the intensity at disc centre by taking the definition of the disc integrated spectrum
        \begin{equation}
            F = \int_{0}^{2\pi}\int_{0}^{R}I_0 f_\mu rdrd\phi
        \end{equation}
        where $I_0=I(\mu=1)$ is the intensity at disc centre, $r$ is the radial variable, $\phi$ is the azimuthal variable, and $R$ is the stellar radius, and make the following change of variables.
        \begin{equation}
            r = R\cos^{-1}{\mu} \,\,\,\,\,,\,\,\,\,\, dr = \frac{R}{-\sin{\frac{r}{R}}}d\mu
        \end{equation}
        This results in an intensity at disc centre of 
        \begin{equation}
            I_0 = \frac{F}{2\pi R^2\int_{0}^{1}f_\mu \mu d\mu}.
        \end{equation}
        
        The final transmission spectrum is generated by applying a transformation to the relative flux spectrum. The transformation is 
        \begin{equation}
            \frac{R_{P,\lambda}}{R_{\star}} = \sqrt{1-\frac{F_\lambda}{F_{\lambda,\star}}}\,,
        \end{equation}
        where $R_{P,\lambda}$ is the wavelength dependent observed planetary radius and $R_{\star}$ is the stellar radius. 
        
\begin{figure*}
        \begin{center}
        	\includegraphics[scale=0.97]{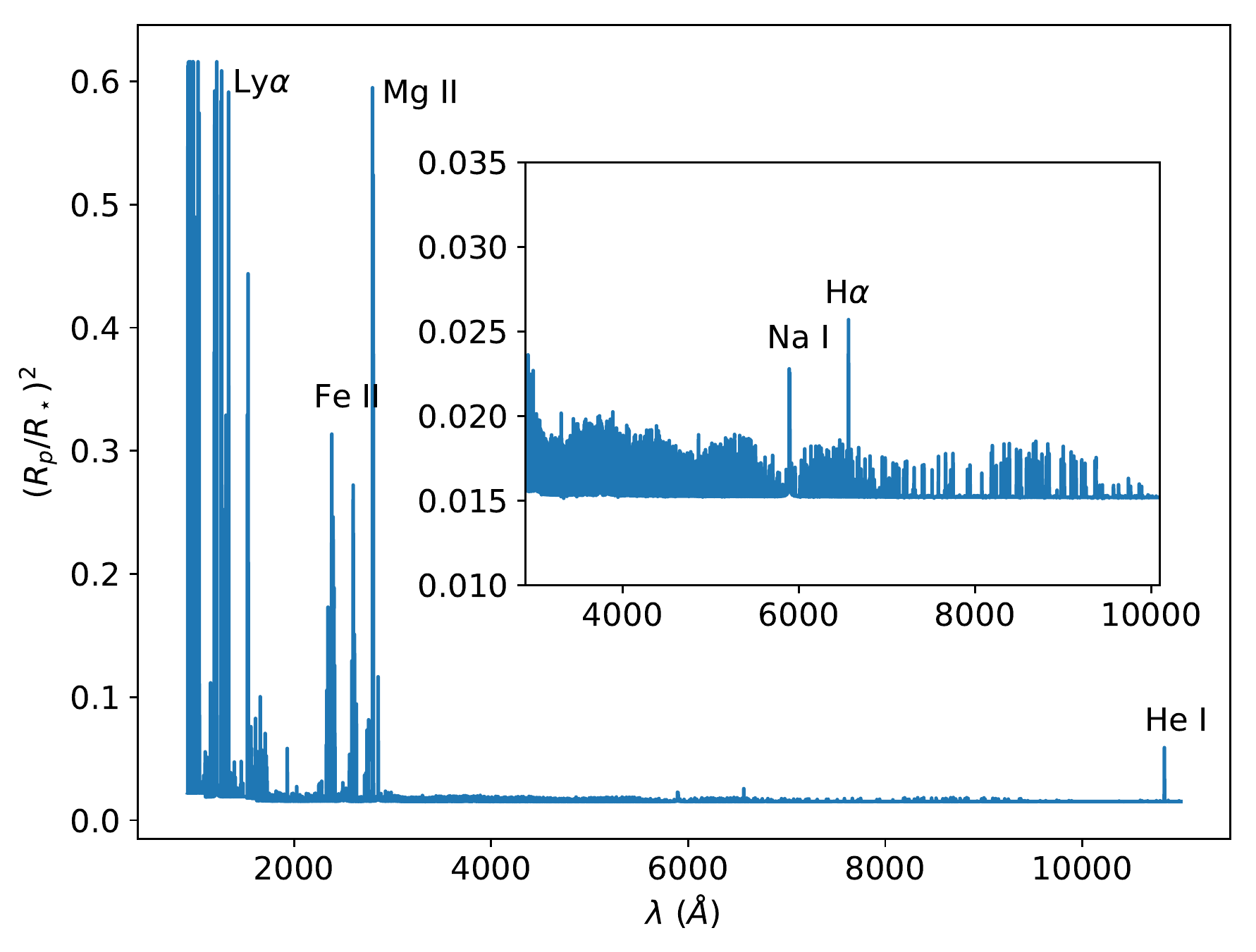}
            \caption{High spectral resolution ($R=100\,000$) synthetic NLTE transmission spectrum of HD\,209458b at mid-transit. Prominent features include Lyman series lines, Fe{\sc ii} packets of lines at $\sim$2350 and 2600 \AA, the Mg{\sc ii}\,h\&k doublet at 2800 \AA, Na{\sc i}\,D lines at 5890 \AA, H$\alpha$, and the He{\sc i} triplet at $\sim$10830 \AA. The inset displays the finer details of the 3000-10000 \AA~range.} 
            \label{fig:trans_spec}
        \end{center}
        \end{figure*}

        Fig.~\ref{fig:trans_spec} presents the resultant synthetic NLTE atomic transmission spectrum of HD\,209458b at mid transit, computed for H, H$_2$, He, Na, Mg, and Fe. The spectrum displays a large number of strong atomic absorption features, several of which extend to very high altitudes, namely the Lyman series and the Mg{\sc ii} $\lambda$2800 \AA\ doublet. There is evidence of strong Fe{\sc ii} absorption around 2350 and 2600 \AA, and a multitude of weak Fe{\sc i} lines spread across the majority of the full spectral range. The He{\sc i} $\lambda$10830 \AA\ triplet is present, as are both H$\alpha$ and the Na\,D doublet as well.
        
\begin{figure}
        	\includegraphics[width=\columnwidth]{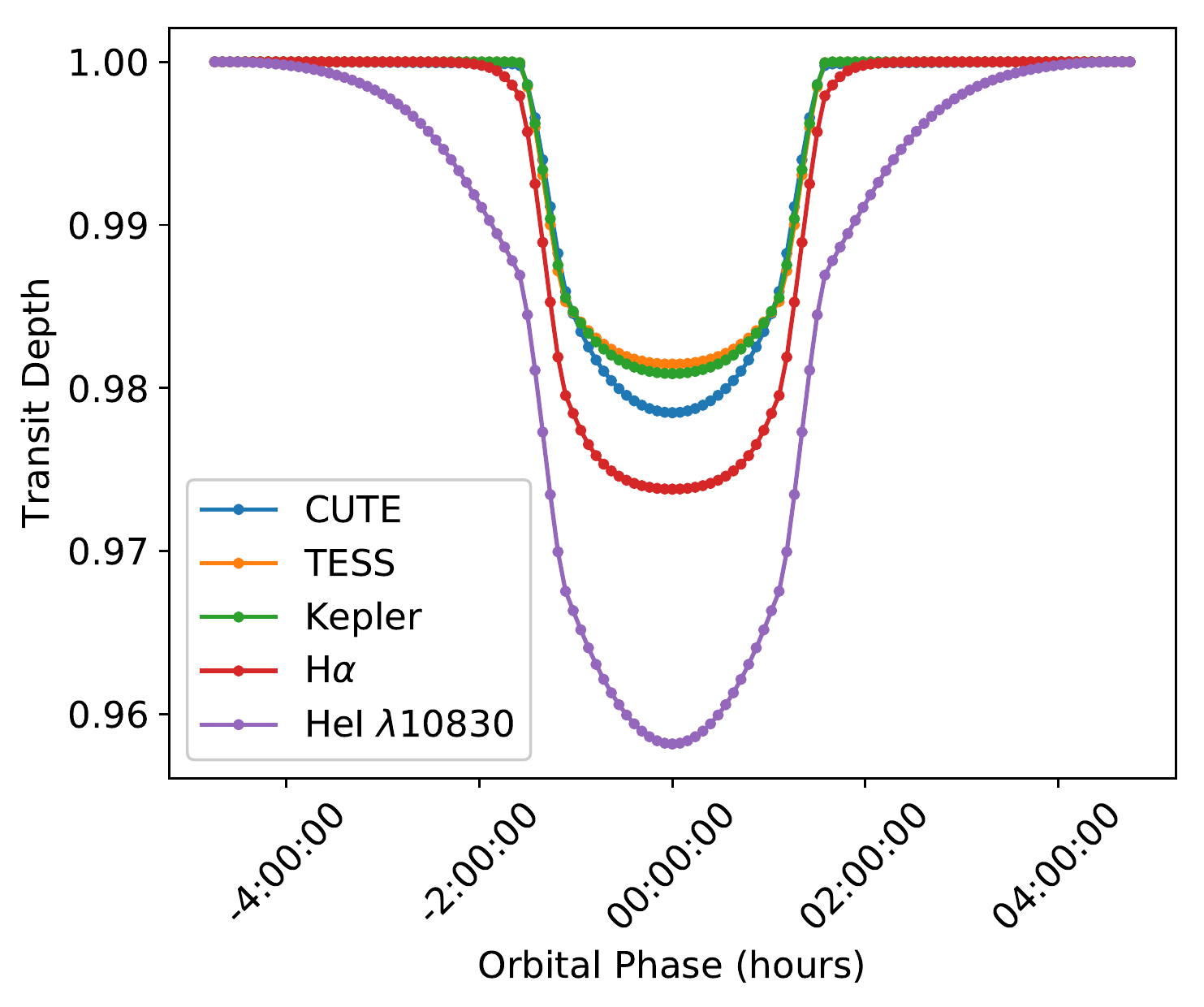}
            \caption{Synthetic NLTE transit light curves in three wide and two narrow wavebands spanning the full synthetic spectral range: CUTE, TESS, Kepler, H$\alpha$, and He{\sc i} $\lambda10830$.}
            \label{fig:light_curve}
        \end{figure}

        Transit light curves are prepared using the same methodology as the transmission spectrum, while changing the value of $D$ by adding an offset perpendicular to the impact parameter such that $D=\sqrt{b^2+x^2}$, where $x$ is the value of the offset. Relative flux spectra are evaluated over a range of values for $x$, simulating the planet transiting the disc of the star, and are then integrated over a given waveband or filter to produce the transit light curve. To demonstrate this process, we have prepared synthetic light curves in a sample of five passbands (three wide and two narrow) spanning the full extent of our synthetic spectrum: the CUTE passband ($\sim2550-3300\,$\AA), the TESS passband ($\sim6000-10000\,$\AA), the Kepler passband ($\sim4350-9000\,$\AA), a narrow H$\alpha$ band ($\lambda_0=6562.8$ \AA, $\delta\lambda\pm2.5$ \AA), and a narrow He{\sc i} $\lambda10830$ band ($\lambda_0=10830$ \AA, $\delta\lambda\pm2.5$ \AA). The synthetic light curves are displayed in Fig.~\ref{fig:light_curve}.
        
\section{Results}
\label{sec:obs}

    In this section, we present the effects of NLTE modelling on our synthetic spectrum, and compare our model spectrum with observed quantities for transit light curves and several strong spectral features. In all cases, synthetic quantities are presented with the same precision as the respective observations for ease of comparison. We confirm the modelled results of \citet{oklopcic18} (He{\sc i} $\lambda10830$ triplet), \citet{fisher19} (Na{\sc i} D doublet), and \citet{barman07} (Fe{\sc i} optical continuum), and demonstrate that NLTE generally increases absorption of spectral features by up to 40\% compared to LTE.
    
    Notably, while the Mg{\sc i} $\lambda2850$ \AA\ and Mg{\sc ii}\,h\&k lines are strong in our synthetic spectrum, recent publications of near-UV (NUV) observations for HD\,209458b have shown that Mg is not detected in the upper atmosphere \citep{cubillos2020}, suggesting that Mg-bearing aerosols are preferentially formed at lower altitudes and that the lower atmosphere is consequently not completely clear. Therefore, we leave Mg out of our comparative analysis here and refer the reader to \citet{cubillos2020} for a thorough discussion on the reasons for a non-detection of Mg in the upper atmosphere of HD\,209458b.
    
    
    

    \subsection{LTE vs NLTE}

\begin{figure}
        	\includegraphics[width=\columnwidth]{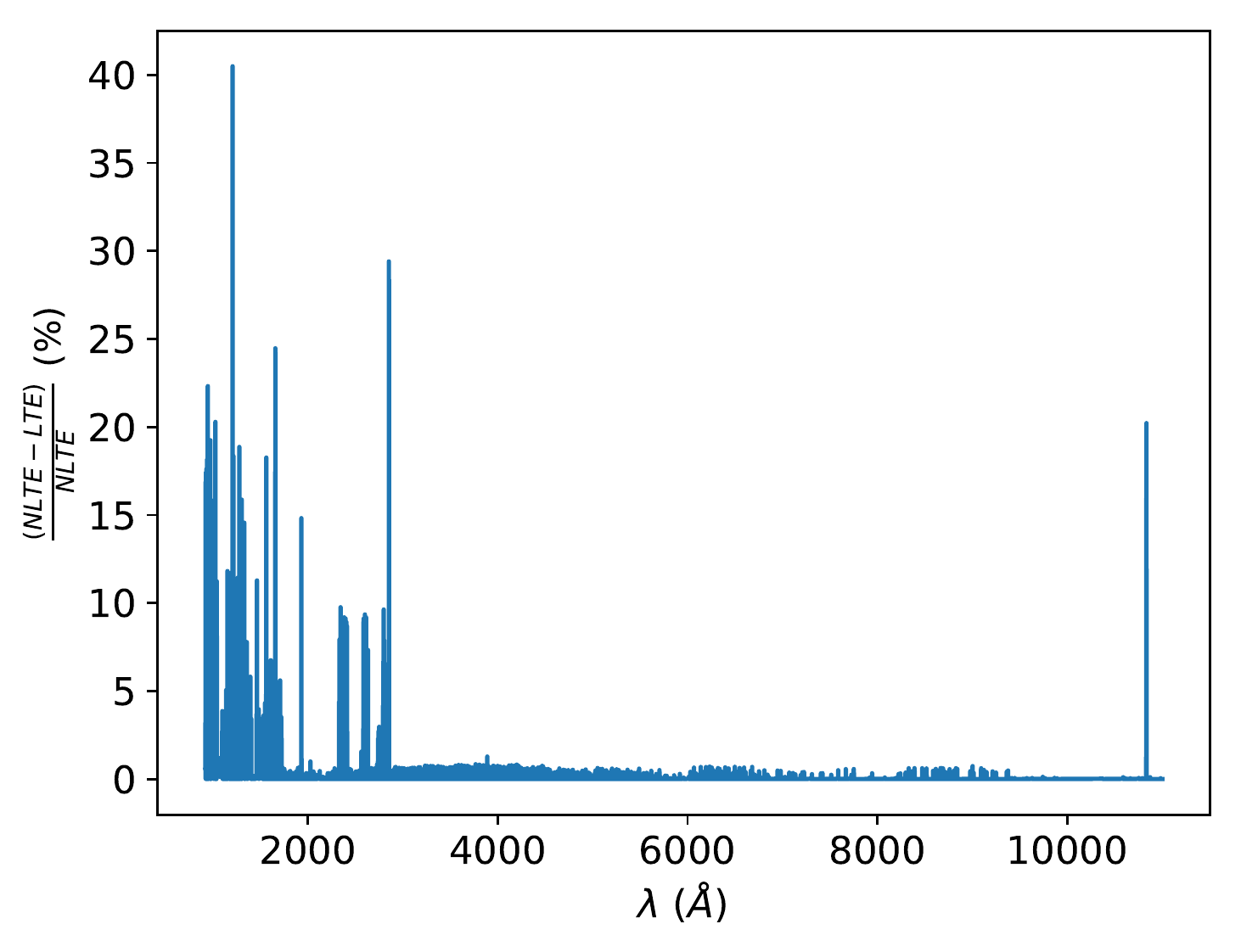}
            \caption{Difference between NLTE and LTE synthetic spectra of HD\,209458b at mid-transit, at high spectral resolution ($R=100\,000$).} 
            \label{fig:spec_compare}
        \end{figure}

        In addition to the NLTE model spectrum, we produced an LTE spectrum following the same methodology described in Section~\ref{sec:trans_spec_mod}. Cloudy's method of LTE spectral synthesis differs from the NLTE synthesis only by using LTE relative level populations rather than detailed level balance equations. All other aspects of the modelling, including photoionization, are identical. We acknowledge that this is not true LTE, given that the ionization isn't calculated according to the Saha equation, but we choose to call this model LTE in work to contrast with the NLTE model. Fig.~\ref{fig:spec_compare} compares our NLTE relative flux spectrum with the LTE one. The average broadband NLTE effect is to increase the transit depth of atomic absorption lines by about 0.3\% while preserving the shape of the lines in the optical and NIR, with larger differences seen for specific lines in the NUV and far-UV (FUV), as well as the He{\sc i} $\lambda10830$ \AA\ feature, with differences ranging between $1$ and $20\%$, depending on the feature. The large NLTE effect for the He{\sc i} triplet is not surprising given that these lines form through a primarily NLTE process.
        
\begin{figure*}
        \begin{center}
        	\includegraphics[scale=0.8]{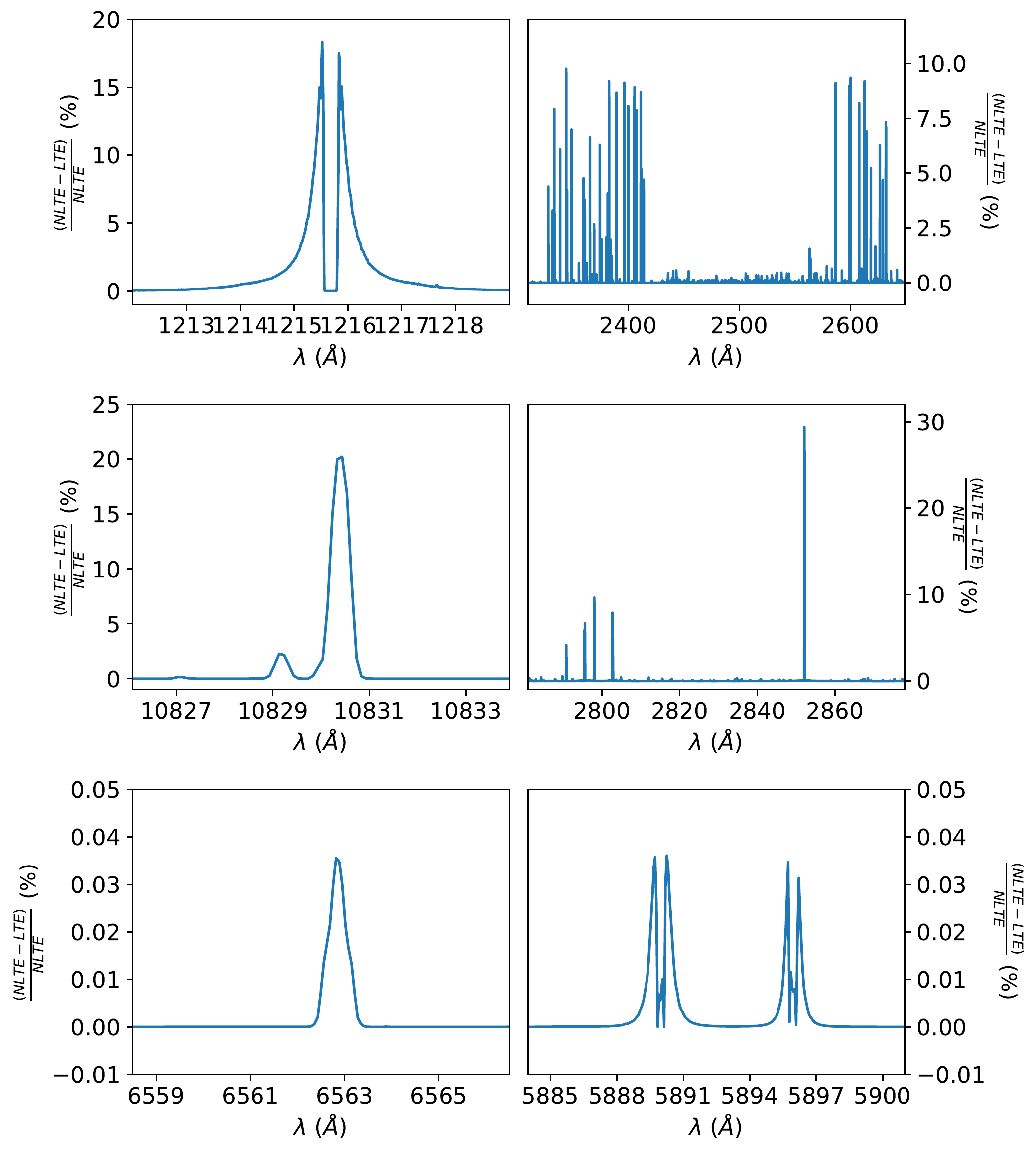}
            \caption{NLTE$-$LTE differences for specific lines and features.  Top-left: Ly$\alpha$. Top-right: Prominent Fe{\sc i} and Fe{\sc ii} bands at NUV wavelengths. Middle-left: He{\sc i} $\lambda10830$ \AA\ triplet. Middle-right: Mg{\sc i} $\lambda$2852 \AA\ line and Mg{\sc ii}\,h\&k lines.  Bottom-left: H$\alpha$. Bottom-right: Na{\sc i}\,D doublet.}
            \label{fig:spec_compare_zoom}
        \end{center}
        \end{figure*}

        The largest NLTE effects, which increase transit depths by $>20\%$, are seen in the Lyman series lines, the Si{\sc iii} $\lambda1206.5$ \AA\, line, the Mg{\sc i} $\lambda2850$ \AA\, line, and the He{\sc i} $\lambda10830$ triplet. 
        The Mg{\sc ii}\,h\&k doublet, as well as the UV Fe{\sc ii} groups of lines centred at $\sim2375$ \AA\, and $\sim2600$ \AA, show differences of $\sim10\%$. The remaining UV lines generally display changes of $5$ to $15\%$, noticeably different from the VIS and NIR lines that show $\leq1\%$. Notably, H$\alpha$ is relatively unaffected by NLTE, as are the Na{\sc i}\,D doublet lines, although the Na lines display a curious behaviour where the wings of the lines show a larger difference than the line cores. \citet{fisher19} likewise found that they were unable to distinguish between the LTE and NLTE line profiles of the Na doublet. 
        The wings of Ly$\alpha$ are stronger in NLTE, but we are unable to comment upon the core of the line: a consequence of how we build our transmission spectrum is that lines are capped at a radius of $R_\star$, and Ly$\alpha$ extends to this limit in both the LTE and NLTE spectra. Regardless, interstellar absorption prevents testing the differences in the core of Ly$\alpha$ for all systems but those with the most extreme radial velocities. Fig.~\ref{fig:spec_compare_zoom} shows the effect of NLTE modelling on these features.
        
    \subsection{Broadband Light Curves}
    \label{sec:light_curves}

        A comparison of our optical light curve transit depths with observed values is presented in Table~\ref{tab:lc_depths}. We find excellent agreement with the HST STIS G750M light curves but it is the only HST configuration that produces a transit depth in agreement with our values. The remaining configurations produce transit depths deeper than our models, by more than $6\sigma$ in the case of the F550W grating. Conversely, our MOST model light curve is deeper than the observations by $12\sigma$.
        
        While we do include H$_2$ Rayleigh scattering in our models, we ignore contributions from other molecular species, and consequently will be underestimating the full effect. The mismatch between the observations and our modelled values may also be a consequence of our assumed planetary and stellar radii, and our planetary pressure-radius calibration. It is also possible that part of the difference can be ascribed also to the use of different limb darkening laws for the data analysis and the computation of the synthetic transit depths.
        
        \begin{table*}
        	\centering
        	\caption{Broadband light curve transit depths at mid transit.}
        	\label{tab:lc_depths}
        	\begin{tabular}{lccccc}
        		\hline
        		\noalign{\smallskip}
        		Instrument & Wavelength Range & No. of & Observed Transit Depth & This Work & Reference\\
        		 & (\AA) & Observations & (\%) & (\%) & \\
        		\noalign{\smallskip}
        		\hline
        		\noalign{\smallskip}
        		MOST & $4000-7000$ & 2 & $1.7526\pm0.0115$ & 1.8949 & 1\\
                HST STIS G750M & $5813-6382$ & 4 & $1.8630\pm0.1289$ & 1.8473 & 2\\
        		HST STIS G750L & \phantom{0}$5240-10270$ & 2 & $2.0215\pm0.0550$ & 1.7916 & 2\\
        		HST STIS G430L & $2900-5700$ & 2 & $2.1290\pm0.0483$ & 1.9583 & 2\\
        		HST FGS F550W & $5100-5875$ & 5 & $2.1662\pm0.0440$ & 1.8764 & 2\\
        		\noalign{\smallskip}
        		\hline
        	\end{tabular}
        	\tablebib{(1)~\citet{rowe08}; (2) \cite{agol07}.}
        \end{table*}

    \subsection{Na{\sc i}\,D doublet}

\begin{figure}
        	\includegraphics[width=\columnwidth]{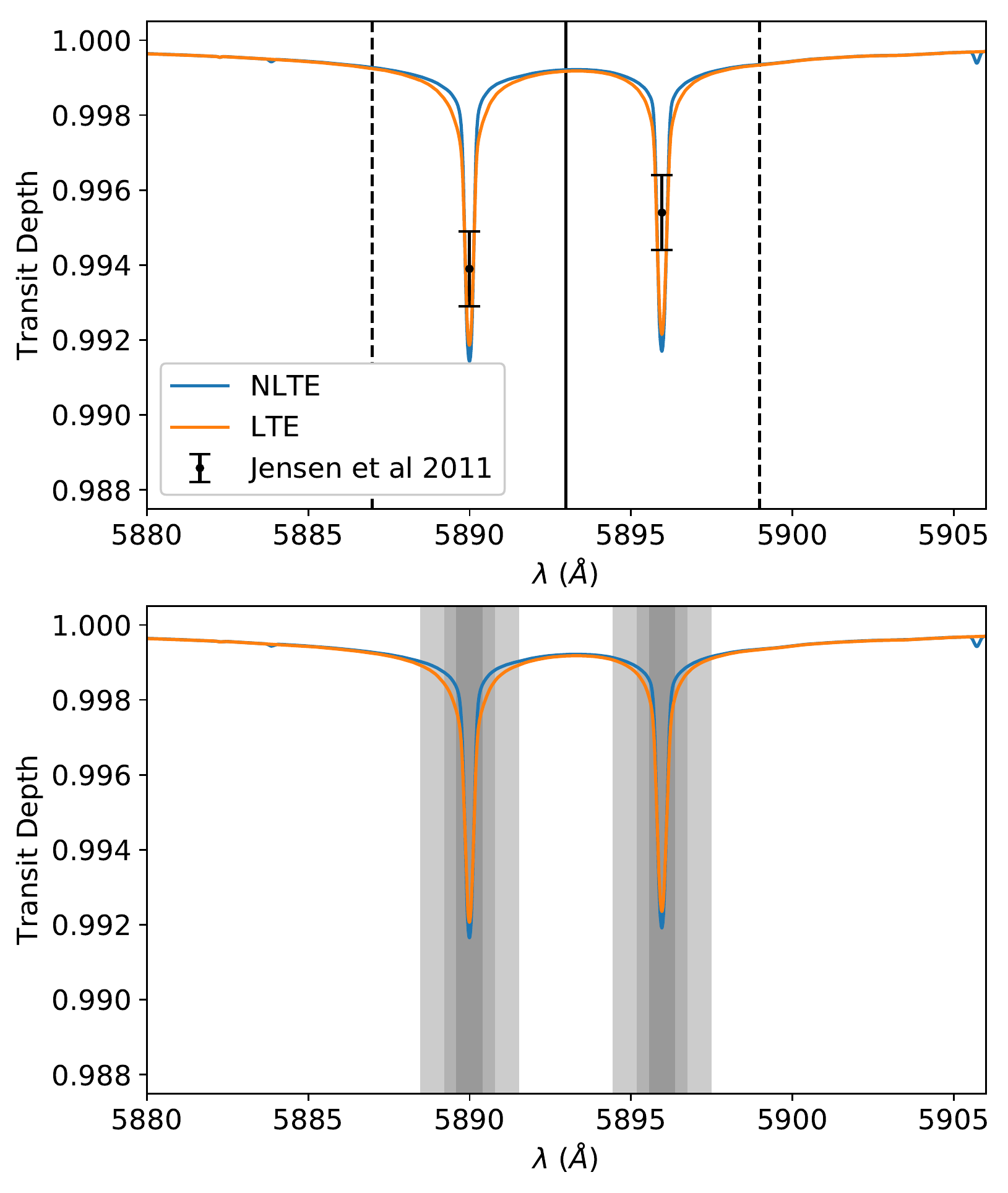}
            \caption{The Na{\sc i}\,D doublet lines in our NLTE and LTE model spectra. {\it Top}: HET HRS comparison spectra ($R=65000$). The solid black line indicates the position centred between the two line cores and the dashed lines indicate the boundaries of the central 12 \AA\ band. The black points indicate the observed core absorption of the lines. {\it Bottom}: Subaru HDS comparison spectra ($R=45\,000$). The shaded areas indicate the narrow (0.75 \AA), medium (1.5 \AA), and wide (3.0 \AA) passbands used in the analysis.}
            \label{fig:NaIDdoublet}
        \end{figure}

        We compare the Na{\sc i}\,D doublet lines in our spectra to those presented in \citet{jensen11} and \citet{snellen08}. A total of 23 in-transit observations of the HD\,209458b Na{\sc i}\,D doublet and 69 out-of-transit observations were obtained by \citet{jensen11} using the Hobby-Eberly Telescope High Resolution Spectrograph (HET HRS; $R=65\,000$), and converted to transmission spectra according to
        \begin{equation}
            S_T = \left(\frac{F_{\rm in}}{F_{\rm out}}\right) -1
        \end{equation}
        where $S_T$ is the transmission spectrum flux normalized to 0, $F_{\rm in}$ is the observed in-transit flux, and $F_{\rm out}$ is the observed out-of-transit flux. They averaged the transmission spectra in three different 12 \AA\ wavebands; a central waveband centered halfway between the cores of the two lines and two additional wavebands immediately to the blue and red of the central waveband. The total absorption level was then calculated as
        
        \begin{equation}
            \label{eqn:absorb}
            M_{\rm abs} = \left<S_T\right>_c - \frac{\left<S_T\right>_b+\left<S_T\right>_r}{2}\,,
        \end{equation}
        
        where $M_{\rm abs}$ is the measured absorption level, and $\left<S_T\right>_c$, $\left<S_T\right>_b$, and $\left<S_T\right>_r$ are the average absorption levels in the central, blue, and red wavebands, respectively. We calculate the absorption of our Na{\sc i}\,D doublet in the same fashion, and present the results in Table~\ref{tab:NaIDjensen}. Additionally, we compare the peak absorption levels in the core of each line.  Fig.~\ref{fig:NaIDdoublet} displays the doublet in our models convolved to the resolution of HET HRS, highlighting the different 12 \AA\ wavebands used in the analysis.
        
        \begin{table*}
        	\centering
        	\caption{Absorption of the Na{\sc i}\,D doublet and comparison with \citet{jensen11}.}
        	\label{tab:NaIDjensen}
        	\begin{tabular}{lcccc}
        		\hline
        		Spectrum & Total Absorption (\%) & $D_1$ peak (\%) & $D_2$ peak (\%) & $\frac{D_1}{D_2}$ absorption ratio\\
        		\hline
        		Observed & $0.0263\pm0.0062$ & $0.61\pm0.10$ & $0.46\pm0.10$ & $1.32\pm0.51$ \\
                NLTE & 0.0993 & 0.86 & 0.83 & 1.04 \\
        		LTE & 0.1127 & 0.81 & 0.78 & 1.04 \\
        		\hline
        	\end{tabular}
        \end{table*}
        
        Our Na absorption is stronger in both our NLTE and LTE spectra than the observed value. The lack of pressure broadened wings in our model spectra artificially increases the absorption when measured in this fashion, which our higher overall absorption can be attributed to. While we find that our NLTE spectrum has deeper line cores than LTE, the LTE spectrum shows a greater overall absorption, attributed to the stronger wings of the doublet lines. Both the NLTE and LTE spectra show a ratio of the $D_1$ to $D_2$ core absorption values that agree within uncertainty with the observed ratio, although our line cores are deeper.
           
        \citet{snellen08} collected 18 in-transit observations of the Na{\sc i}\,D doublet observed with the Subaru High Dispersion Spectrograph (HDS; $R=45\,000$). They measured the depth of the Na{\sc i}\,D features in the transmission spectrum within three spectral passbands, with widths of 0.75 \AA, 1.5 \AA, and 3.0 \AA\ centered on each of the lines, and calculated the total absorption level as the central waveband minus the average of the blue and red wavebands, similar to above, and average the results of the separate lines. We again calculate the absorption of our Na{\sc i}\,D doublet in the same fashion, and present the results in Table~\ref{tab:NaIDsnellen}. Fig.~\ref{fig:NaIDdoublet} displays the doublet convolved to the resolution of Subaru's HDS. Our derived absorption values are again stronger than those observed. This agrees with the previous analysis using 12 \AA\ bands, where the models showed stronger absorption than observed, although the difference is not as large in these narrow bands. 
        
        \begin{table*}
        	\centering
        	\caption{Absorption of the Na{\sc i}\,D doublet and comparison with \citet{snellen08}.}
        	\label{tab:NaIDsnellen}
        	\begin{tabular}{lccc}
        		\hline
        		Spectrum & 0.75 \AA~band (\%) & 1.5 \AA~band (\%) & 3.0 \AA~band (\%) \\
        		\hline
        		Observed & $0.135\pm0.017$ & $0.070\pm0.011$ & $0.056\pm0.007$ \\
                NLTE & 0.320 & 0.195 & 0.120 \\
        		LTE & 0.312 & 0.215 & 0.138 \\
        		\hline
        	\end{tabular}
        \end{table*}

    \subsection{H$\alpha$}

        The HET HRS transmission spectroscopy observations of HD\,209458b centered at the H$\alpha$ line display an odd behaviour \citep{jensen12}. While the authors did not detect an overall absorption or ``emission'' signal over a 16 \AA\ band centered on the stellar line core, they note that the average transmission spectrum shows a dramatic feature with a ``spike'' to the blue of the stellar line center and a ``dip'' to the red. Both features peaked at roughly 0.5\% absolute deviation from zero and are several angstroms wide. The feature was also roughly symmetric when reflected about the zero point and line center. While they do not report their signal-to-noise ratio for their observations, we note that our H$\alpha$ lines both exhibit greater than $1\%$ absorption in their cores, and would be observable with similar observations.
        
        We measure the absorption of our NLTE and LTE H$\alpha$ lines according to Eq.~(\ref{eqn:absorb}), using 16 \AA\ bands, and get absorption measurements of $M_{abs,NLTE}=0.0281\%$ and $M_{abs,LTE}=0.0300\%$. We note that there are several additional lines within the 16 \AA\ bands that are present in the NLTE spectrum but not the LTE spectrum, and caution the reader that differences in the measured absorption is not uniquely caused by NLTE effects on H$\alpha$ alone. A narrower 2~\AA\ band reveals absorption measurements of $M_{abs,NLTE}=0.278\%$ and $M_{abs,LTE}=0.249\%$, confirming that the stronger LTE absorption in the 16~\AA\ band is caused by additional lines in the NLTE spectrum, and not limited to H$\alpha$.
        While we acknowledge the possibility that Cloudy lacks some relevant physics, one should also consider the possibility of problems in the data reduction and/or analysis that led to the odd shape of the transmission spectrum. Therefore, further observations and/or an independent re-analysis of the available data would be very valuable for providing additional constraints to the atmospheric properties of HD\,209458b.

    \subsection{He{\sc i} $\lambda10830$ \AA\ triplet}

\begin{figure}
        	\includegraphics[width=\columnwidth]{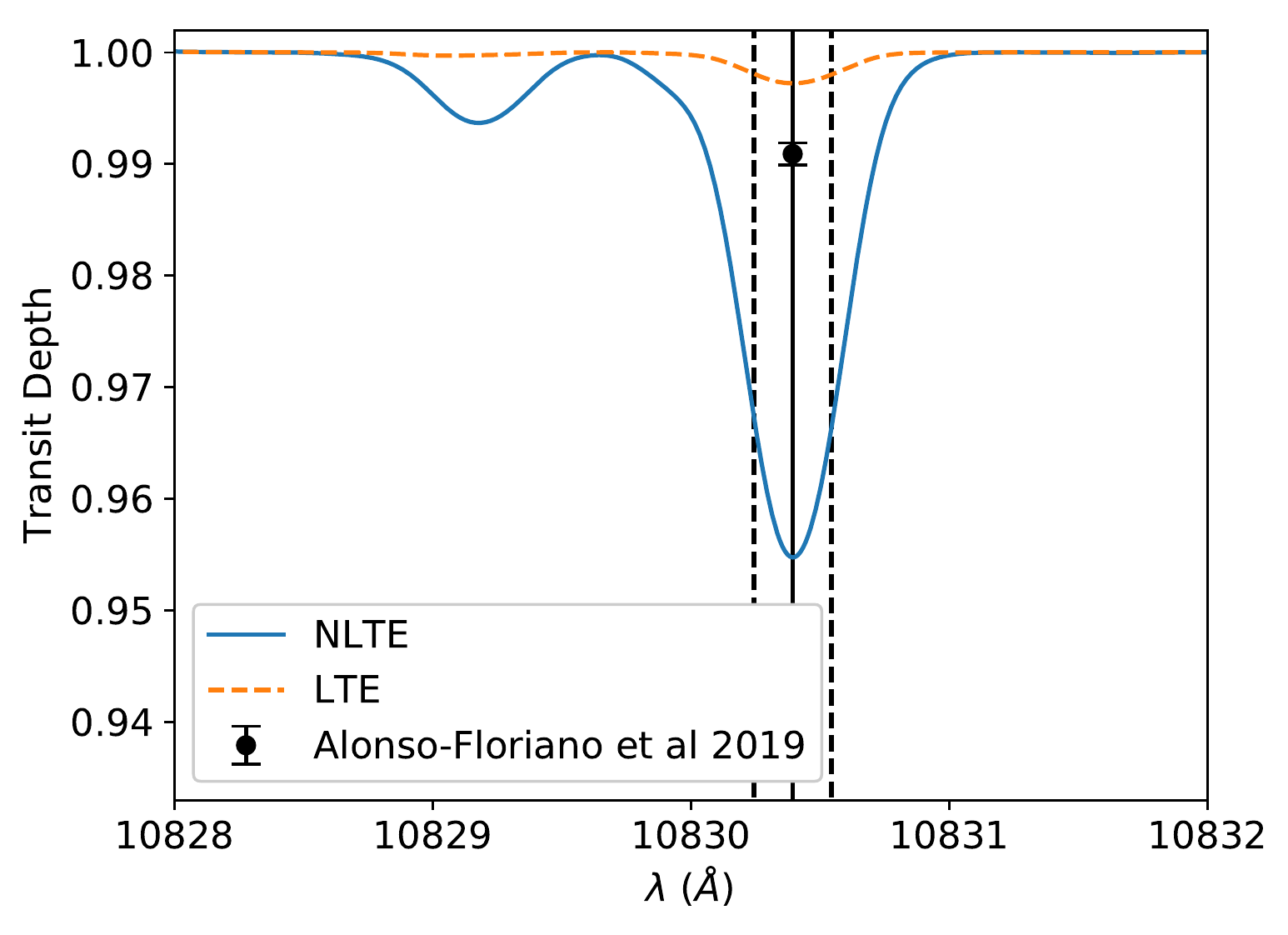}
            \caption{The He{\sc i} $\lambda$10830 triplet lines in our NLTE and LTE model spectra. The solid black line indicates the position of the core of the strong blended feature, the dashed lines indicate a 0.3~\AA\ band centred on the line core, and the black point indicates the observed level of core absorption.}
            \label{fig:He10830}
        \end{figure}

        \citet{alonso-floriano19} collected 33 observations of the HD\,209458b He{\sc i} IR triplet in transmission with the CARMENES NIR channel. They took two measures of the strength of the planetary absorption, the peak value of the average absorption signal in the core of the two strongest and blended lines and the average value over a $0.3$ \AA\ bandwidth centred on the core of the feature. We take the same measures of both our NLTE and LTE spectra, after convolving them to the spectral resolution of CARMENES NIR ($R=80\,400$), and present the results in Table~\ref{tab:HeI10830}. Our NLTE and LTE line profiles are displayed in Fig.~\ref{fig:He10830}. 
        
        \begin{table*}
        	\centering
        	\caption{Absorption of the He{\sc i} $\lambda10830$ \AA~triplet}
        	\label{tab:HeI10830}
        	\begin{tabular}{lccc}
        		\hline
        		Spectrum & Core Absorption (\%) & $0.3$ \AA~Band (\%) & $\frac{\rm Core}{\rm Band}$  ratio\\
        		\hline
        		Observed & $0.91\pm0.10$ & $0.71\pm0.06$ & $1.28\pm0.24$ \\
                NLTE & 4.52 & 2.85 & 1.59 \\
        		LTE & 0.28 & 0.18 & 1.56 \\
        		\hline
        	\end{tabular}
        \end{table*}
        
        Neither the NLTE nor the LTE values agree with the observations for either measure, with the NLTE model showing stronger absorption than observed by $\sim3.6\%$, and the LTE showing weaker absorption by $\sim0.6\%$. We additionally investigate the ratio of the core absorption to the absorption over the $0.3$ \AA~band, and find that our models both exhibit ratios larger than the observed value, suggesting narrower line profiles than observed. This indicates that either the 1-D atmospheric model temperature profile is too cool at the altitudes where He{\sc i} absorption is optically thick, or that we are missing and additional source of broadening, such as microturbulence, in our modelling. 
        
        In \citet{oklopcic18}, the authors develop a simple 1D model of an escaping atmosphere comprised of only H and He, and predict the absorption of the He{\sc i} $\lambda10830$ triplet lines. They assume the planet transits the center of the stellar disc, and do not include limb darkening in their model. For an HD\,209458b-like planet under these assumptions, they predict depths in the line cores of 0.44\% and 2.38\% for the weak and strong lines respectively, and a combined equivalent width (EW) of the feature of 0.014~\AA. Using the same set of assumptions to prepare a NLTE transmission spectrum with Cloudy, we find depths of 0.38\% and 2.55\%, and an EW of 0.012~\AA, in line with the predictions of \citet{oklopcic18}. The agreement between our model and that of \citet{oklopcic18}, which describes in detail the excitation chemistry, is an important benchmark. This agreement gives us confidence in the capability of Cloudy to model theoretical planetary atmospheres and transmission spectra.
        

    \subsection{Fe{\sc i} and Fe{\sc ii}}

\begin{figure}
        	\includegraphics[width=\columnwidth]{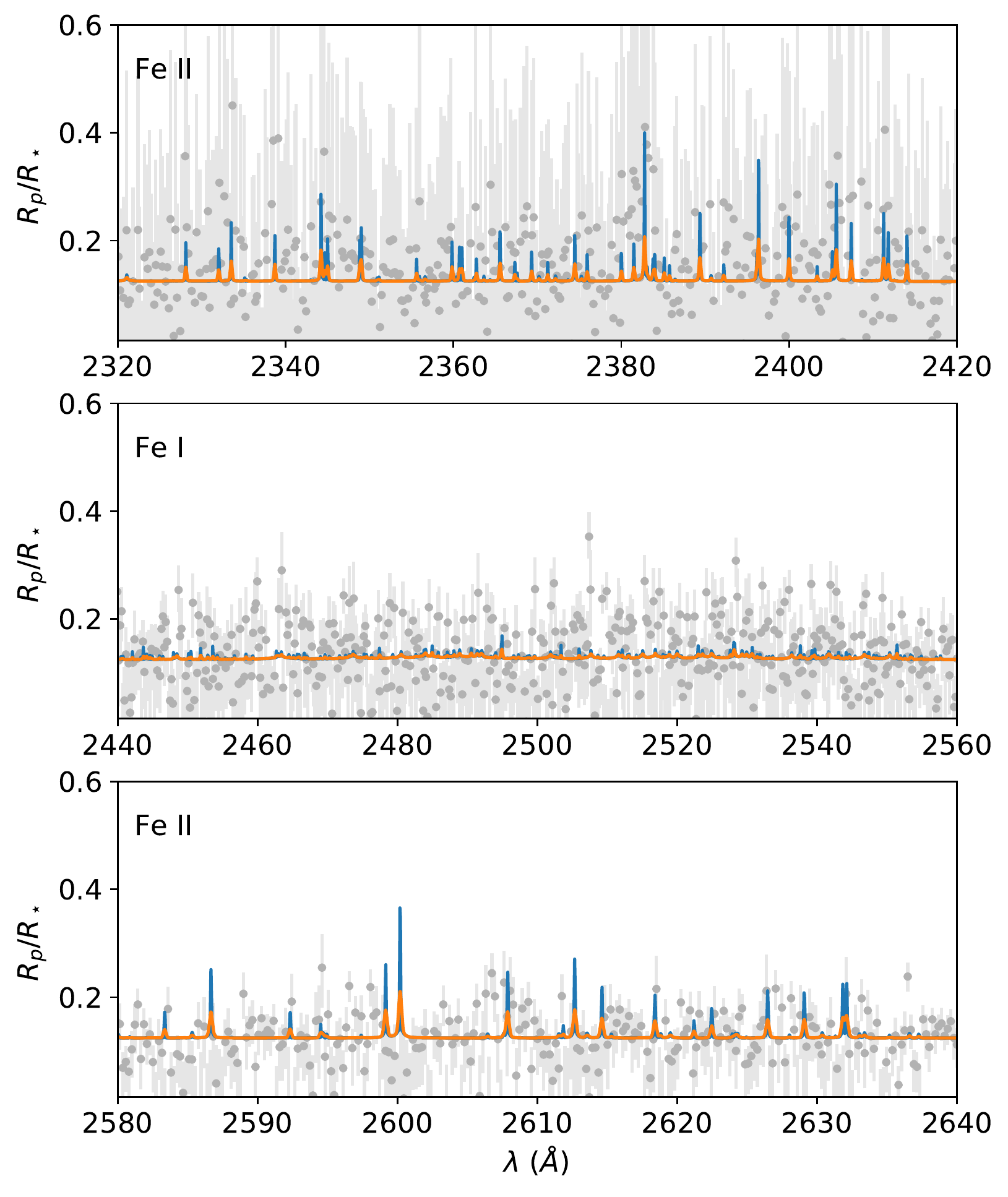}
            \caption{NUV Fe{\sc i \& ii} bands. The grey points with error bars denote the observed spectrum at a resolution of $R\sim12\,000$, the blue curve is our high-resolution NLTE spectrum, and the orange curve is our spectrum convolved to the resolution of the observed data. The top and bottom panels show bands of prominent Fe{\sc ii} lines, and the middle panel shows a band of Fe{\sc i} lines.}
            \label{fig:Fe_bands}
        \end{figure}

        We compare our model spectra to three archival NUV Fe-band transmission observations of HD209458b, obtained with HST STIS \citep{vidal-madjar13} using the E230M grating. The data consists of \'echelle spectra comprising 23 orders, in which each order consists of 1024 wavelength samples, with a spectral resolution of $R=30\,000$. The entire STIS spectrum covers the $2300-3100$ \AA\ range, with some overlap between the orders. We treat the synthetic spectra as in \citet{cubillos2020}, re-sampling the spectrum at a resolution of $R=12\,000$. We find that our NLTE spectrum generally provides a better fit to the observations, thus we focus on the NLTE comparison here. Fig.~\ref{fig:Fe_bands} displays the observed spectra, overlayed with our NLTE model spectrum at both the Cloudy output high resolution and convolved to the observed resolution.
        
        We note that our model fits the strength of the Fe{\sc ii} band centered at $2600$ \AA~more accurately than the one centred at $2375$ \AA; our models show similar line strengths in the two bands, while the observed data shows stronger absorption in the $2375$ \AA~band. This effect of the two bands having different line strengths has also been observed to occur in the NUV transmission spectrum of WASP-121b \citep{sing19}. This difference is unexpected because the relative strength of the two bands is controlled by atomic line parameters, the quality of which has been confirmed by comparisons with spectra of well studied stars \citep{landstreet2011}. One possible explanation for why our model produces this behavior is that Cloudy uses super-levels to compute NLTE level populations for ions more complicated than the H-like and He-like iso-electronic sequences. Super-levels are the higher energy levels in model atoms grouped together into single or a few representative levels, usually one per principle quantum number. Fe{\sc i} and {\sc ii} are both extremely complicated atoms to model for detailed NLTE calculations, but current model atoms are being constantly improved \citep[e.g.,][]{bautista17}, making the use of super-levels a simplifying approximation rather than a necessity.

\begin{figure}
        	\includegraphics[width=\columnwidth]{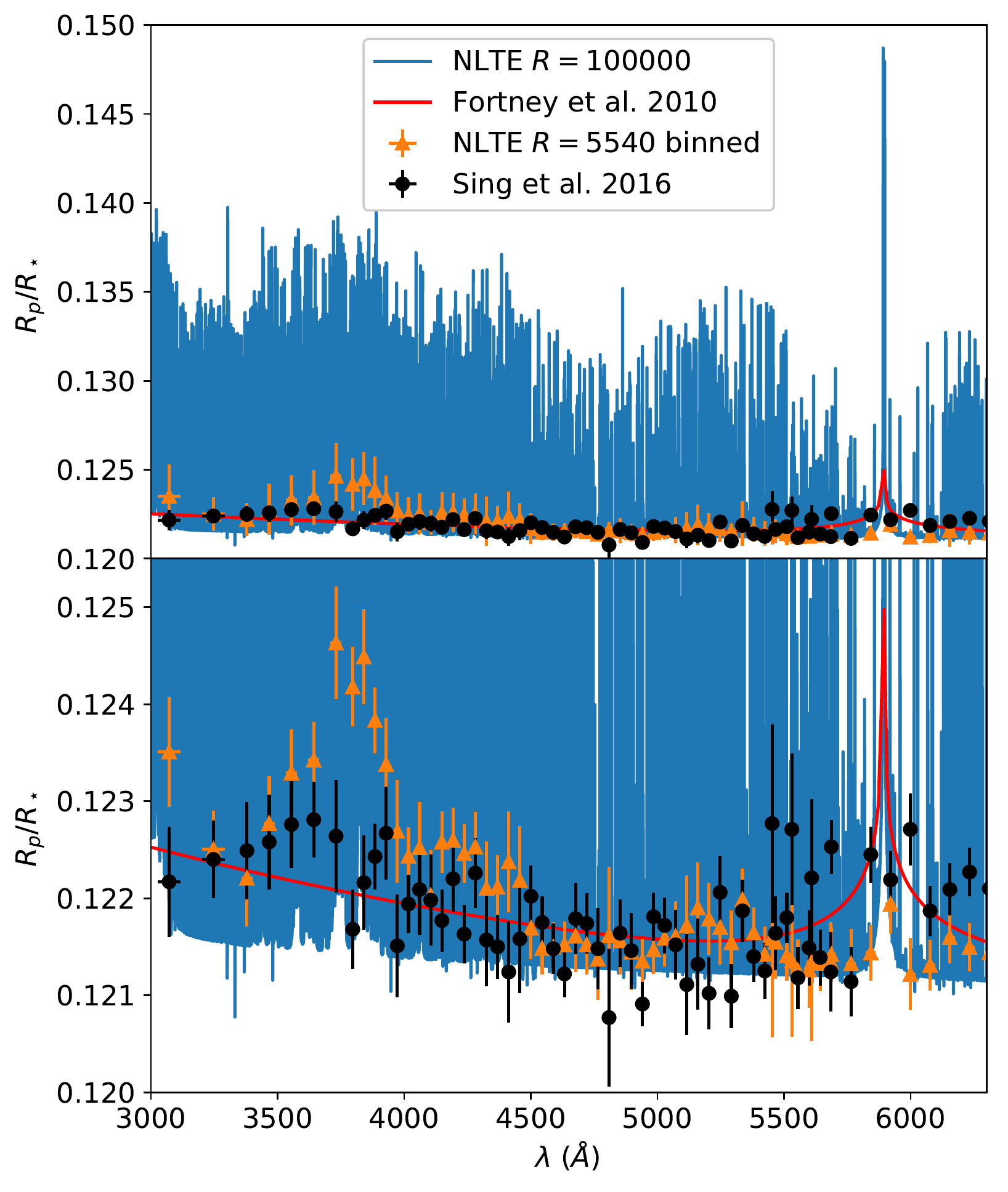}
            \caption{Comparison of the optical continuum slope between our NLTE spectrum and observations. $Top:$ Full region, displaying strength of high resolution atomic features. $Bottom:$ Zoomed view of the continuum level in this region.}
            \label{fig:fe_slope}
        \end{figure}

        \citet{barman07} demonstrated that the slope of the optical continuum in the range of $\lambda\approx3000$ to $6000$ \AA~could be reproduced at low spectral resolution $(R\sim5000)$ in a cloudless atmosphere, as a result of metallic line blanketing, primarily Fe{\sc i} lines. We compare our NLTE spectrum with HST STIS G750M, G750L, and G430L observations of HD~209458b \citep{sing16}, and the LTE spectral model of \citet{fortney10}. The observations were made at a spectral resolution of $R\sim5540$, and binned with a variable bin size depending on the STIS grating. Our high resolution spectrum displays stronger continuum absorption than either the observed data or model, which can be attributed to differences in the values used for the planetary and stellar radii. We consequently match our continuum level to the modelled level at $9300$ \AA, shifting our spectrum down by $\Delta R_p/R_\star=-0.00223$. Fig.~\ref{fig:fe_slope} displays our NLTE spectrum comparison with the model and observations. We find that the continuum level in the high resolution spectrum is indeed weaker than both the predicted and observed levels, but when convolved to the spectral resolution of and binned in the same manner as the observations, our synthetic spectrum that accounts for both H$_2$ Rayleigh scattering and Fe{\sc i} line blanketing is able to reproduce the slope of the continuum in this region, with the exception of two features. 
        
        As discussed in Section~\ref{sec:trans_spec_mod}, Cloudy does not include pressure broadening of lines in its spectral synthesis, and the wings of our Na{\sc i} D doublet are much weaker than observed. Additionally, we display a strong broadband absorption feature centred at $\sim 3750$ \AA\ not present in the observations. While the placement of the feature would suggest that it is related to the Ca {\sc ii} H and K lines, this is coincidental as our model does not include Ca. The feature is the result of blending many atomic absorption lines, the majority of which are Fe{\sc i}. When considering the Fe{\sc i} and Fe{\sc ii} features in the NUV and optical together, it appears that an atmosphere in which Fe is more ionized than our model below the 1\,$\mu$bar level may be a better fit to the data. In fact, this would lead to weaker absorption around 3750 \AA\ and stronger absorption at the position of the Fe{\sc ii} NUV bands. 
        
    \subsection{C, O, and Si UV Lines}

        Far-ultraviolet observations of HD~209458b have been performed with both the HST STIS G140L low resolution grating ($R\sim600$) \citep{vidal-madjar04}, and the HST COS G130M high resolution grating ($R\sim17000$ to $20000$) \citep{linsky10, ballester15}. While there is some debate over the observability and detection of a number of FUV diagnostic lines, we choose four to focus on: Si{\sc iii} $\lambda1206$, O{\sc i} $\lambda1302$, C{\sc ii} $\lambda1335$, and Si{\sc ii} $\lambda1527$. 

        \citet{vidal-madjar04} fit the observed transit light curves in narrow windows ($\sim10$ \AA) centred on the target lines, leaving $\left(R_{p,\lambda}/R_\star\right)^2$ as a free parameter in their fit, recording the fitted value as their measure of the absorption for these lines. Table~\ref{tab:UV_lines_1} compares their values with the values derived in a similar manner from our NLTE and LTE spectra. While they present strong detections of the C{\sc ii} and O{\sc i} features, both of their Si features are consistent with non-detections. In these cases, they reported the $2\sigma$ upper limits on the absorption depths. Our measured values agree with the observations and upper limits in all cases except for O{\sc i} $\lambda1302$, where we find less absorption than the observations by greater than $2\sigma$; however we warn the reader that these comparisons do not account for interstellar medium absorption affecting the shortest wavelength lines of the O{\sc i} triplet and C{\sc ii} doublet.
        
        \begin{table*}
        	\centering
        	\caption{HST STIS G140L low resolution absorption depths for UV spectral lines, measured as $\left(R_{p,\lambda}/R_\star\right)^2$ in a narrow window around the lines. NLTE and LTE values computed in a similar fashion for comparison.}
        	\label{tab:UV_lines_1}
        	\begin{tabular}{lcccc}
        		\hline
        		Spectral Feature & Window (\AA) & Vidal-Madjar 2004 (\%) & NLTE (\%) & LTE (\%) \\
        		\hline
                C{\sc ii} $\lambda1335$ & 1332-1340 & $7.5^{+3.6}_{-3.4}$ & 5.2 & 5.1 \\
                O{\sc i} $\lambda1302$ & 1300-1310 & $12.8\pm4.5$ & 3.2 & 2.7 \\
                Si{\sc ii} $\lambda1527$ & 1525-1536 & $<47.4$ & 3.0 & 2.7 \\
                Si{\sc iii} $\lambda1206$ & 1204-1210 & $<5.9$ & 3.6 & 4.0 \\
        		\hline
        	\end{tabular}
        \end{table*}
        
        Both \citet{linsky10} and \citet{ballester15} measure the absorption as the average ratio of the in- to out-of-transit fluxes in a $\pm50$ km\,s$^{-1}$ window centred on the lines, although \citet{ballester15} claim that the \citet{linsky10} results are invalid for not recording the stellar flux at the time of their observations, and including errors for photon noise only, ignoring the intra-transit variations in stellar flux. Results from both are presented in Table~\ref{tab:UV_lines_2}, alongside absorption values computed in a similar fashion from our NLTE and LTE spectra. While the \citet{ballester15} results show stronger absorption for the C{\sc ii} $\lambda1335$ doublet, their Si{\sc iii} $\lambda1206$ absorption is consistent with a non-detection, and both measurements exhibit uncertainties greater than those of \citet{linsky10} by factors of $\sim6.5$. Our measured values agree with all the observed values with the exception of Si{\sc iii} $\lambda1206$ in our NLTE spectrum being weaker than the result reported by \citet{linsky10}. The comparison for the  C{\sc ii} $\lambda1335$ doublet has to be taken with care because our transmission spectra do not account for interstellar medium absorption. When considering just the line at longer wavelength, which is not affected by the interstellar medium, we obtain an absorption of 11.0\% in NLTE and 11.1\% in LTE, while the observed absorption is 7.9$\pm$1.5\% \citep{linsky10}.
        
        \begin{table*}
        	\centering
        	\caption{HST COS G130M high resolution absorption depths for UV spectral lines, measured as the ratio of in to out of transit flux in a $\pm50$ km\,s$^{-1}$ centred on the lines. NLTE and LTE values computed in a similar fashion for comparison.}
        	\label{tab:UV_lines_2}
        	\begin{tabular}{lcccc}
        		\hline
        		Spectral Feature & Linsky 2010 (\%) & Ballester 2015 (\%) & NLTE (\%) & LTE (\%) \\
        		\hline
                C{\sc ii} $\lambda1335$ & $7.8\pm1.3$ & $9.3\pm8.4$ & 8.9 & 9.1 \\
                Si{\sc iii} $\lambda1206$ & $8.2\pm1.4$ & $8.6\pm9.5$ & 6.0 & 7.9 \\
        		\hline
        	\end{tabular}
        \end{table*}   
   
\section{Conclusions}

    We have presented a generalized framework for producing synthetic transmission spectra of exoplanets, given a planetary $T$--$P$ profile as input, using the NLTE spectral synthesis hydrostatic code Cloudy. The framework is designed to work for a broad range of exoplanetary atmospheric conditions and to produce transmission spectra at a desired spectral resolution in a given waveband, with the possibility to account for stellar limb darkening. To demonstrate our framework, we generated both an LTE and NLTE transmission spectrum for the well studied exoplanet HD\,209458b starting from a published planetary $T$--$P$ profile that accounts for hydrodynamics. 
    
    We presented a differential comparison of the NLTE and LTE spectra, highlighting prominent features with strong NLTE effects. We also performed comparisons with observed broadband transit light curves, the Na{\sc i} D doublet, H$\alpha$, the He{\sc i} $\lambda10830$ triplet, Fe{\sc i} and Fe{\sc ii} bands in both the UV and optical, and several FUV lines including Si{\sc iii} $\lambda1206$, the O{\sc i} $\lambda1302$ triplet, the C{\sc ii} $\lambda1335$ doublet, and Si{\sc ii} $\lambda 1527$.
    
    Our simulations match one out of five observed broadband transit depths and the mismatch may be caused by one or a combination of the following: the lack of molecular absorption and aerosols in the transmission spectra; differences in the assumed planetary and stellar radii; the adopted planetary pressure-radius calibration; differences in the considered limb darkening laws. Both our NLTE and LTE spectra show stronger total absorption than observed at the position of the Na{\sc i}\,D lines, but the comparison is biased by the fact that our model does not account for pressure broadening. Our measured absorption values for the FUV lines agree with observations for three of the four lines, but are weaker than the observed O{\sc i} $\lambda1302$ absorption.
    
    We were unable to compare our synthetic H$\alpha$ absorption with observations because of the odd behaviour present in the observed spectrum, but note that the strength of our line should be observable, given the noise level of the observations. We suggest that further H$\alpha$ observations and/or an independent re-analysis of the existing ones should be carried out to clarify the presence and possible strength of this feature. 
    
    Neither our NLTE nor our LTE spectra match the observed He{\sc i} $\lambda10830$ absorption, with our NLTE model overestimating the absorption by $\sim3.6\%$, and our LTE model underestimating the absorption by $\sim0.6\%$. Additionally, we find that our line profiles are narrower than the observed lines, suggesting either that the temperature profile in the atmospheric model is too cool in the region where He{\sc i} becomes optically thick, or that the excitation chemistry needs to be revised, or that we are missing an additional source of broadening such as microturbulence, or a combination of those. We were able to reproduce the predicted absorption of He{\sc i} $\lambda10830$ under the same set of assumptions as \citet{oklopcic18}.
    
    We find differences up to the 10\% level between our LTE and NLTE spectra of the NUV Fe{\sc i \& ii} bands, with the NLTE spectrum being a better fit to the data. Both LTE and NLTE models fit the Fe{\sc ii} $\lambda2600$ band better than the $\lambda2375$ band, where we underestimate the absorption. We confirm the result of \citet{barman07} that the slope of the optical continuum can be reproduced in low resolution spectra by Fe{\sc i} line blanketing and H$_2$ Rayleigh scattering, without additional absorption. We propose that additional Fe ionization below the 1 $\mu$bar level in our model would provide better fits to the data in both the optical and UV.
    
    A differential comparison of our model spectra revealed that NLTE modelling generally increases the absorption of features by $\leq1\%$ in the optical and NIR, and $5$ to $20\%$ in the UV, but can increase absorption by up to $\sim40\%$ for individual features (e.g. the Mg{\sc i} $\lambda2850$ or Si{\sc iii} $\lambda1206.5$ lines). We found negligible difference between LTE and NLTE for both H$\alpha$ and the Na{\sc i} D doublet. We have benchmarked this application of Cloudy by reproducing the the results of \citet{oklopcic18}, \citet{fisher19}, and \citet{barman07} for He{\sc i} $\lambda10830$, Na{\sc i} D, and optical Fe{\sc i} continuum absorption, respectively. We conclude that accounting for NLTE effects is necessary for modelling planetary spectral lines forming in the upper atmosphere, particularly at UV wavelengths.
    
    The modelling framework is open to several improvements. Currently, the inclusion of any of the following would improve the physical realism of the model: hydrodynamics, molecular chemistry, pressure broadening, 3-D atmospheric structure, stellar activity, and Doppler effects. While Cloudy lacks the ability to include hydrodynamics, pressure broadening, and 3-D structure in our framework, it is possible to include stellar activity and Doppler effects by making several additional assumptions and knowledge of additional planetary and stellar parameters.
    
    
    


\begin{acknowledgements}
    This research has made use of the Extrasolar Planet Encyclopaedia, the Exoplanet Transit Database \citep{poddany10}, NASA's Astrophysics Data System Bibliographic Services, the NASA Exoplanet Archive, which is operated by the California Institute of Technology, under contract with the National Aeronautics and Space Administration under the Exoplanet Exploration Program, and the SVO Filter Profile Service (http://svo2.cab.inta-csic.es/theory/fps/) supported from the Spanish MINECO through grant AYA2017-84089. This research also made use of the NIST Atomic Spectra Database funded [in part] by NIST's Standard Reference Data Program (SRDP) and by NIST's Systems Integration for Manufacturing Applications (SIMA) Program. MEY acknowledges funding from the \"OAW-Innovationsfonds IF\_2017\_03.
    
    The following software and packages were used in this work: \texttt{Cloudy v17.01} (\citealt{ferland17}); \texttt{Python v2.7}; \texttt{Python} packages \texttt{Astropy} \citep{astropy13}, \texttt{NumPy} \citep{numpy06,numpy11}, and \texttt{Matplotlib} \citep{matplotlib07}.
\end{acknowledgements}

\bibliographystyle{bibtex/aa} 
\bibliography{ms} 

\end{document}